\documentclass[prd,aps,floats,amssymb,reprint,nofootinbib]{revtex4}
\pdfoutput=1

\usepackage{amsmath}
\usepackage{amsfonts}
\usepackage{amssymb}
\usepackage{latexsym}
\usepackage{graphicx}
\usepackage{slashed}
\usepackage{subfig}

\newcommand{\order}[1]{\mathcal{O}(#1)}
\newcommand{\units}[1]{\mathrm{#1}}
\newcommand{\vev}{\mathsf{v}}
\def\Yeq{Y_\textrm{EQ}}

\begin{document}


\title{Dark Matter with Density-Dependent Interactions}
\author{Kimberly K. Boddy$^{(1)}$\footnote{kboddy@theory.caltech.edu},
Sean M. Carroll$^{(1)}$\footnote{seancarroll@gmail.com},
and Mark Trodden$^{(2)}$\footnote{trodden@physics.upenn.edu}}

\affiliation{$^{(1)}$California Institute of Technology, Pasadena, CA 91125, USA. \\
$^{(2)}$Center for Particle Cosmology, Department of Physics and Astronomy, University of Pennsylvania, Philadelphia, PA 19104, USA.}

\date{28 December 2012}

\begin{abstract}
The decay and annihilation cross sections of dark matter particles may depend on the value of a chameleonic scalar field that both evolves cosmologically and takes different values depending on the local matter density. This possibility introduces a separation between the physics relevant for freeze-out and that responsible for dynamics and detection in the late universe. We investigate how such dark sector interactions might be implemented in a particle physics Lagrangian and consider how current and upcoming observations and experiments bound such dark matter candidates. A specific simple model allows for an increase in the annihilation cross section by a factor of $10^6$ between freeze-out and today, while more complicated models should also allow for scattering cross sections near the astrophysical bounds.
\end{abstract}

\maketitle

\setcounter{footnote}{0}

\section{Introduction}

The particle physics properties of dark matter are important for three distinct aspects of its behavior: they determine how the initial abundance of dark matter arose, they govern how the dark matter distribution evolves and influences structure formation, and they delineate the possible ways in which dark matter may be detected. Of course, these three roles are not typically independent, since they all depend on the prescribed interactions between the dark matter particles themselves and also between dark matter and the Standard Model. These connections often provide a powerful motivation for particular dark matter candidates -- for example, the freeze-out abundance of weakly interacting massive particles points to new physics at the weak scale, which in turn leads to an attractive connection between dark matter and proposed solutions to the hierarchy problem, such as weak-scale supersymmetry.

The idea that dark matter could have interactions of astrophysically interesting magnitude has received a good amount of attention \cite{Carlson:1992fn,deLaix:1995vi,Spergel:1999mh,Wandelt:2000ad,Firmani:2000ce,Rocha:2012,Peter:2012}, motivated in part by purported discrepancies between the standard $\Lambda$CDM model and observations of structure on small scales (as described in \cite{Salucci:2007tm}, for example). While most approaches of this form concentrate on giving an appreciable scattering cross-section to the dark matter, it is also interesting to consider enhanced annihilation cross sections \cite{Kaplinghat:2000vt}.

One obstacle to simple implementations of this idea is that the required cross section for a thermal relic to obtain the right relic abundance is close to the weak scale, far too small to be relevant to dynamics in the late universe. In this paper we explore the idea that the dark matter cross section might be much larger now than it was at freeze-out, due to the evolution of a background field.

In a cosmological context, the evolution of background fields can assert a significant influence on the properties of dark matter as a function of spatial location or cosmic epoch \cite{Casas:1991ky,Anderson:1997un,Amendola:1999er,Hoffman:2003ru,Farrar:2003uw,Bean:2007nx,Bean:2007ny,Bean:2008ac,Corasaniti:2008kx,Cohen:2008nb}. A straightforward way to achieve such effects is to invoke a light scalar field that interacts with dark matter and/or ordinary matter as well as through its own potential, and whose expectation value feeds into the dark-matter properties. A popular scenario along these lines is the ``chameleon mechanism,'' which acts to screen light, cosmologically relevant degrees of freedom to protect them from precision local tests of gravity~\cite{Khoury:2003aq,Khoury:2003rn,Brax:2010kv,Gannouji:2010fc,Mota:2010uy}.

In this paper we investigate dark matter that interacts through a gauge symmetry with a coupling constant that depends on a chameleonlike scalar field. (The effects of chameleon vector bosons on laboratory experiments were considered in \cite{Nelson:2008tn}.) Just as the properties of a cosmologically relevant scalar can be drastically modified in the presence of local density inhomogeneities or after evolving over cosmic time, so the interactions of dark matter may be  modified. We are able to find a model in which the late-time interaction strength is considerably higher than that at freeze-out -- although admittedly, this behavior does not seem generic.

We begin by reexamining the conventional story of dark matter freeze-out according to the Boltzmann equation, but with the additional ingredient that the dark matter properties are evolving with time. We then look at specific models featuring a Dirac dark matter particle and a U(1) gauge symmetry that is spontaneously broken, along with a chameleon scalar field. We study the cosmological evolution of this coupled system and calculate the dark matter properties, including annihilation and scattering cross sections. Finally we exhibit numerical solutions to a specific model, showing that the annihilation cross section can increase substantially during cosmic evolution.

\section{The General Picture: Evolving Dark Matter in the Early Universe}
Before discussing specific models, let us first consider how the usual story of dark matter freeze-out might be modified if the annihilation cross section depends on the dynamics of another field. In the next section, we will explore Lagrangians that couple the dark matter to a scalar field that affects its interaction cross sections. For simplicity we work in a flat Friedmann-Robertson-Walker (FRW) universe, described by the metric $ ds^2 = -dt^2 + a^2(t) \times\left(dx^2 + dy^2 + dz^2 \right)$, with scale factor $a(t)$.

The decoupling of dark matter takes place in the early universe in the radiation-dominated regime, in which particles with masses $m\ll T$ are the dominant component of the cosmic energy budget. To a good approximation, we may therefore ignore contributions from nonrelativistic species in thermal equilibrium with the radiation and approximate the energy density as
\begin{equation}
  \rho_\textrm{R} = \frac{\pi^2}{30} g_* T^4
\end{equation}
and the entropy density as
\begin{equation}
  s = \frac{2\pi^2}{45} g_{*S} T^3 \ ,
\end{equation}
where, as usual,
\begin{align}
  g_* &= \sum_{i=\textrm{bosons}} g_i \left(\frac{T_i}{T}\right)^4
  + \frac{7}{8} \sum_{i=\textrm{fermions}} g_i \left(\frac{T_i}{T}\right)^4 \\
  g_{*S} &= \sum_{i=\textrm{bosons}} g_i \left(\frac{T_i}{T}\right)^3
  + \frac{7}{8} \sum_{i=\textrm{fermions}} g_i \left(\frac{T_i}{T}\right)^3
\end{align}
and $g_i$ is the number of internal degrees of freedom for particle species $i$.

For $T \gtrsim 300~\units{GeV}$, $g_{*S}=g_*=106.75$, which includes all particles in the Standard Model. When $100~\units{MeV} \gtrsim T \gtrsim 1~\units{MeV}$, the electron and positron are relativistic and so $g_{*S}=g_*=10.75$. At the temperature of the CMB today, $T_0=2.725~\units{K}$, $g_{*S,0}=3.91$, and $g_{*,0}=3.36$.

Consider a dark sector that was in thermal equilibrium with the visible sector at some very high temperature scale, below which they decouple effectively enough to consider each sector separately to be in equilibrium. The visible sector is at temperature $T$ with entropy density $s(T)$, while the dark sector is at temperature $T_d$ with entropy density $s_d(T_d)$. The expansion of the universe is governed by both sectors with
\begin{equation}
  g_*^\textrm{tot}(T) = g_*(T) + g^d_* (T_d) \left(\frac{T_d}{T}\right)^4 \ ,
\end{equation}
but quantities in the dark sector (for example, the dark matter annihilation cross section and number density) are determined by $T_d$~\cite{Feng:2008mu}.

Since the entropy in each sector is conserved independently, the assumption that the two sectors were in equilibrium at some unification scale at time $t_u$ allows us to express the dark bath temperature in terms of the visible bath temperature at some later time $t$ via
\begin{equation}
  \frac{g^d_{*S}(t)}{g_{*S}(t)} \frac{T_d^3 (t)}{T^3 (t)}
  = \frac{g^d_{*S}(t_u)}{g_{*S}(t_u)} \ .
\end{equation}
All Standard Model particles contribute at $t_u$ to give $g_{*S}(t_u)=106.75$, and all dark particles contribute to $g^d_{*S}(t_u)$. In what follows, we will use the temperature of the visible sector and convert $T_d$ to $T$ as needed. For convenience we write
\begin{equation}
  \xi (t) = \frac{T_d (t)}{T(t)} = \left(\frac{g_{*S}(t)}{g_{*S}^d(t)}
  \frac{g_{*S}^d(t_u)}{g_{*S}(t_u)}\right)^{1/3} \ .
\end{equation}

The success of big bang nucleosynthesis (BBN) and the structure of the cosmic microwave background (CMB) power spectrum place tight bounds on any new relativistic degrees of freedom in the dark sector. The limit on the effective number of light neutrino species is $N_\nu = 3.24 \pm 1.2$ at the 95\% confidence level~\cite{Cyburt:2004yc}, which gives
\begin{equation}
  g_*^d \xi^4(t_\textrm{BBN}) = \frac{7}{8} \times 2 \times (N_\nu-3) \leq 2.52
  \quad \mbox{(95\% confidence)}
  \label{eq:neutrino-bounds}
\end{equation}
for 3 light SM neutrino species~\cite{Ackerman:2008gi}. The 5-year WMAP data ~\cite{Hinshaw:2008kr} also bounds the number of neutrino species by $N_\nu = 4.4 \pm 1.5$ at the 65\% confidence level, and the 7-year WMAP data ~\cite{Larson:2010gs} places a tighter lower limit of $N_\nu > 2.7$ at the 95\% confidence level.

\subsection{The Boltzmann Equation}

Let us assume the dark matter $\psi$ is a stable particle that annihilates with a thermalized annihilation cross section $\left<\sigma v\right>$. The general Boltzmann equation governing the number density $n$ of a particle of mass $m$ is
\begin{equation}
  \dot{n} + 3H n + \left<\sigma v\right> (n^2 - n_\textrm{EQ}^2) = 0 \ ,
\end{equation}
where $H$ is the Hubble parameter
\begin{equation}
  H = \frac{\dot{a}}{a} = \sqrt{\frac{8}{3}\pi G \rho_R}
    = \sqrt{\frac{4\pi^3 Gg_*^\textrm{tot}}{45}}\; T^2
\end{equation}
and $n_\textrm{EQ}$ is the equilibrium number density
\begin{equation}
  n_\textrm{EQ} \approx \frac{g}{(2\pi)^3} \int d^3\vec{p}\, e^{-E/T_d}
               = \frac{g}{2\pi^2} m^2 \xi T K_2\left(\frac{m}{\xi T}\right) \ ,
  \label{eq:neq}
\end{equation}
where $K_2$ is the modified Bessel function of the second kind of order two. Generalizing the traditional treatment, we allow for the possibility that the mass of the dark matter $\tilde{m}_\psi(\phi)$ is a function of a real scalar chameleon field $\phi$ and denote $\phi$-dependent masses and couplings with a tilde.

It is convenient to scale out the effects of the expansion of the universe by defining
\begin{equation}
  Y \equiv \frac{n_\psi}{s}
\end{equation}
($n_\psi(x)$ and $Y(x)$ are taken to be independent of $\phi$) and to use a new independent variable, related to the cosmic time $t$ through
\begin{equation}
  x(t) \equiv \frac{m_T}{T(t)} \ ,
\end{equation}
where $m_T$ is some constant mass scale. In the usual derivation, $m_T$ is chosen to coincide with the dark matter mass; however, since our dark matter has varying mass, we use this constant parameter instead. Defining
\begin{equation}
  b = \sqrt{\frac{45}{4\pi^3 G}}\frac{1}{m_T}
\end{equation}
allows us to write
\begin{equation}
  \frac{dx}{dt} = \frac{m_T}{bx} \sqrt{g^\textrm{tot}_*} \ ,
  \label{eq:xdot}
\end{equation}
which can be used to rewrite the Boltzmann equation for the dark matter as
\begin{equation}
  Y'(x) + \frac{B}{x^2}(Y^2 - \Yeq^2) = 0 \ .
  \label{eq:Boltzmann-eq}
\end{equation}
Here a prime denotes a derivative with respect to $x$, and
\begin{equation}
  B = \left<\sigma v\right> \frac{2\pi^2}{45}
  \frac{g_{*S}}{\sqrt{g^\textrm{tot}_*}}bm_T^2 \ ,
\end{equation}
which may depend implicitly on $\phi$ in our model via a $\phi$ dependence in the cross section. Note that, in terms of these new variables, the equilibrium term is
\begin{equation}
  \Yeq = \frac{45g}{(2\pi^2)^2g_{*S}}
  \left(x\frac{\tilde{m}_\psi(\phi)}{m_T}\right)^2
  \xi K_2\left(\frac{x}{\xi}\frac{\tilde{m}_\psi(\phi)}{m_T}\right) \ ,
\end{equation}
with $g=2$ for Dirac dark matter.

It remains, at this level, to specify $Y(x_i)$, the initial condition for $Y$. We consider $\Delta \equiv Y-\Yeq$, the departure from equilibrium~\cite{K&T}, which obeys
\begin{equation}
  \Delta' = -\Yeq' - \frac{B}{x^2} \Delta (2\Yeq + \Delta) \ .
\end{equation}
At early times ($1 < x \ll x_f$), $Y$ tracks $\Yeq$ extremely closely such that $\Delta$ and $|\Delta'|$ are small. Note that in the non-relativistic approximation, $T \ll \tilde{m}_\psi(\phi)$,
\begin{equation}
  \Yeq \sim x^{3/2} e^{-(x/\xi)(\tilde{m}_\psi(\phi)/m_T)} \ ,
\end{equation}
and so $\Yeq' / \Yeq \approx - \tilde{m}_\psi(\phi)/\xi m_T$ and $\Delta' \approx 0$. Thus, the required initial condition is
\begin{equation}
  Y(x_i) = \Yeq(x_i) + \frac{x_i^2 \tilde{m}_\psi(\phi_i)}{2B\xi m_T} \ ,
  \label{eq:IC-Y}
\end{equation}
where $B(\phi_i)$ and $\tilde{m}_\psi(\phi_i)$ are evaluated at the initial value $\phi_i=\phi(x_i)$.

After the freeze-out value $x_f$, $Y(x)$ will asymptotically approach a constant value $Y_\infty$. The energy density of non-relativistic dark matter today is then
\begin{align}
  \rho_0 &= \tilde{m}_\psi(\phi_0) n_\psi(x_0)
          = \tilde{m}_\psi(\phi_0) Y_\infty s_0 \nonumber\\
         &= \tilde{m}_\psi(\phi_0) Y_\infty \frac{2\pi^2}{45} g_{*S,0} T_0^3 \ .
\end{align}

Having generalized the usual treatment of dark matter as a fluid to the case in which there is a chameleon field determining the dark matter properties, we now turn to specific examples of particle physics models in which these phenomena might arise.
\section{Gauged Dark Matter}
Consider dark matter to consist of a Dirac fermion $\psi$, charged under a dark U(1) gauge group with gauge boson $A_\mu$, and a dark Higgs field $\Phi$ that spontaneously breaks the U(1). We also introduce a chameleonlike field $\phi$ that is a real scalar field with properties that depend on the dark matter energy density. The chameleon couples to the other particles in the dark sector by entering into the dark matter mass $\tilde{m}_\psi(\phi)$, the U(1) coupling $\tilde{f}(\phi)$, and other couplings described below. We consider only an isolated dark sector so that we may investigate the properties of this simple model without the complications of coupling to the visible sector.

\subsection{A Toy Model for Varying Coupling}
As a first step, let us consider the QED Lagrangian with a real scalar field $\phi$, but in which we allow the coupling constant $e$ to vary as a function of spacetime~\cite{Steinhardt:2003iu}.  Specifically, it can vary as a function of $\phi$. Let us write the new coupling as $\tilde{f}(\phi)$. Thus,
\begin{equation}
  \mathcal{L}_{\textrm{QED}\phi} =
  -\frac{1}{2} \partial_\mu\phi \partial^\mu\phi - V(\phi)
  -\frac{1}{4\tilde{f}^2(\phi)} F^{\mu\nu}F_{\mu\nu}
  + i\bar{\psi} \slashed{\partial} \psi
  - m_\psi\bar{\psi}\psi - \bar{\psi} \gamma^\mu \psi A_\mu  \ ,
\end{equation}
where $F_{\mu\nu}=\partial_{\mu}A_{\nu}-\partial_{\nu}A_{\mu}$. Making the redefinition $A_\mu \to \tilde{f}(\phi) A_\mu$, we obtain
\begin{equation}
   \mathcal{L}_{\textrm{QED}\phi} =
   -\frac{1}{2} \partial_\mu\phi \partial^\mu\phi - V(\phi)
   + i\bar{\psi} \slashed{\partial} \psi
  - m_\psi\bar{\psi}\psi - \tilde{f}(\phi) \bar{\psi} \gamma^\mu \psi A_\mu
  - \frac{1}{4 \tilde{f}^2} \left[\partial^\mu (\tilde{f} A^\nu)
    - \partial^\nu (\tilde{f} A^\mu) \right]^2 \ .
\end{equation}
Both Lagrangians are equivalent, but now the gauge transformation reads
\begin{align}
  \tilde{f}(\phi) A_\mu &\to \tilde{f}(\phi) A_\mu + \partial_\mu\omega \\
  \psi &\to e^{-i\omega} \psi \\
  \bar{\psi} &\to e^{+i\omega} \bar{\psi} \ .
\end{align}
If we can neglect factors of ($\partial_\mu \tilde{f} / \tilde{f}$) compared to all other mass scales in the theory (except the Planck mass), then the Lagrangian simplifies to the approximately gauge-invariant form
\begin{equation}
  \mathcal{L}_{\textrm{QED}\phi} \approx
  -\frac{1}{2} \partial_\mu\phi \partial^\mu\phi - V(\phi)
  -\frac{1}{4} F^{\mu\nu}F_{\mu\nu} + i\bar{\psi} \slashed{\partial} \psi
  - m_\psi\bar{\psi}\psi - \tilde{f}(\phi) \bar{\psi} \slashed{A} \psi
\end{equation}
with U(1) current
\begin{equation}
  j^\mu (x) = \tilde{f}(\phi) \bar{\psi} \gamma^\mu \psi \ .
\end{equation}

\subsection{The Cosmological Equations of Motion}
We now include gravity and a complex dark Higgs field $\Phi$ to break the U(1) symmetry and give the dark gauge field a mass. We allow for a varying dark matter mass by using the effective mass parameter $\tilde{m}_\psi(\phi)$, and in the spirit of effective field theory, we also allow all couplings [not just the U(1) coupling $\tilde{f}(\phi)$] to depend on $\phi$.

Neglecting factors of ($\partial_\mu \tilde{f} / \tilde{f}$), the action is then
\begin{align}
  S \approx \int d^4x \sqrt{-g}
  & \left[\frac{\mathcal{R}}{16\pi G}
    -\frac{1}{2}g^{\mu\nu}\nabla_\mu\phi \nabla_\nu\phi - V(\phi)
    - (D_\mu \Phi)^\dagger D^\mu \Phi - V_0(\Phi) \right. \nonumber\\
    & \left. {}- \frac{1}{4} F^{\mu\nu} F_{\mu\nu}
    + i\bar{\psi} \slashed{D} \psi - \tilde{m}_\psi(\phi) \bar{\psi} \psi
    - \tilde{\lambda}_\psi(\phi) (\Phi+\Phi^\dagger) \bar{\psi} \psi \right] \ ,
  \label{eq:fullaction-dirac}
\end{align}
where the gauge covariant derivative is $D_\mu = \nabla_\mu + i \tilde{f}(\phi) A_\mu$. The equations of motion for the fields then follow as
\begin{align}
  \left(i\slashed{D} - \tilde{m}_\psi(\phi) - \tilde{\lambda}_\psi(\phi)
       (\Phi+\Phi^\dagger) \right) \psi &=0\\
   \Box\phi - V'(\phi) - \tilde{m}_\psi^\prime(\phi) \bar{\psi}\psi
       - \tilde{f}'(\phi) \bar{\psi} \slashed{A} \psi
       - \tilde{\lambda}'_\psi(\phi) (\Phi+\Phi^\dagger) \bar{\psi} \psi &=0 \ ,
\end{align}
where a prime denotes differentiation with respect to $\phi$. Let us assume that the universe is dark-charge symmetric, so the average charge current density is negligible compared to the dark matter number density [see \eqref{eq:psibarpsi} below]. Thus, the term proportional to $\tilde{f}' / \tilde{f}$ should be small compared to the one containing $\tilde{m}' / \tilde{m}$, given that $\tilde{f}' / \tilde{f} \sim \tilde{m}' / \tilde{m}$ to within a few orders of magnitude -- a condition we will enforce later. We may write this last equation as
\begin{equation}
  \Box\phi - V'(\phi) - \tilde{m}_\psi^\prime(\phi) \bar{\psi}\psi
  - \tilde{\lambda}'_\psi(\phi) (\Phi+\Phi^\dagger) \bar{\psi} \psi \approx 0 \ .
\end{equation}

We will arrange for the dark Higgs to have a sufficiently large mass that its perturbations are negligible and simply replace $\Phi$ by $\left<\Phi\right>$ in the equations of motion.  The VEV generates an additional mass term for $\psi$, but we can redefine $\tilde{m}_\psi(\phi)$ to absorb this term. We then have
\begin{subequations}
\begin{align}
  \left(i\slashed{D} - \tilde{m}_\psi(\phi) \right)\psi & \approx 0 \\
  \Box\phi - V'(\phi) - \tilde{m}_\psi^\prime(\phi)\bar{\psi}\psi & \approx 0 \ .
  \label{eq:phi-eom0-dirac}
\end{align}
\end{subequations}

We calculate the energy-momentum tensor for $\psi$ by varying the action with respect to the metric. Taking care to correctly handle the nontrivial metric dependence of the covariant derivative~\cite{Birrell-GR}, we have
\begin{equation}
  T_{\mu\nu}^{(\psi)} = -\frac{i}{2} \left[\bar{\psi} \gamma_{(\mu} \nabla_{\nu)} \psi
    - (\nabla_{(\mu} \bar{\psi}) \gamma_{\nu)} \psi  \right]
  + \tilde{f}(\phi) \bar{\psi} \gamma_{(\mu} A_{\nu)} \psi \ ,
\end{equation}
where we have integrated by parts and used the field equation of motion. Taking the trace, we obtain
\begin{align}
  g^{\mu\nu} T_{\mu\nu}^{(\psi)}
  &= -\frac{i}{2} \left[\bar{\psi} \slashed{\nabla} \psi
    - \bar{\psi} \overleftarrow{\slashed{\nabla}} \psi \right]
  + \tilde{f}(\phi) \bar{\psi} \slashed{A} \psi \nonumber \\
  &= -\frac{1}{2} \left[\bar{\psi} i(\slashed{\nabla}
    +i \tilde{f}(\phi)\slashed{A}) \psi
    - \bar{\psi} i(\overleftarrow{\slashed{\nabla}}
    -i \tilde{f}(\phi)\slashed{A}) \psi \right]  \nonumber \\
  &= -\tilde{m}_\psi(\phi) \bar{\psi}\psi \ ,
\end{align}
where, again, we have used the Dirac equation for $\psi$ and $\bar{\psi}$ to obtain the last line. If we model the dark matter as nonrelativistic dust, its pressure is zero and so the trace of the stress tensor is approximately given by $-\rho_\psi$. Thus,
\begin{equation}
  \rho_\psi = \tilde{m}_\psi(\phi) \bar{\psi}\psi \ .
  \label{eq:psibarpsi}
\end{equation}
As a final step in this section, we use this result to rewrite the $\phi$ equation of motion~\eqref{eq:phi-eom0-dirac} as
\begin{equation}
   \Box\phi - V_\textrm{eff}'(\phi) = 0 \ ,
   \label{eq:phi-eom-dirac}
\end{equation}
where the effective potential is
\begin{align}
  V_\textrm{eff}
  &= V(\phi) + \tilde{m}_\psi(\phi) n_\psi \nonumber\\
  &= V(\phi) + \tilde{m}_\psi(\phi) Y(x) \frac{2\pi^2}{45} g_{*S}
  \left(\frac{m_T}{x}\right)^3 \ .
  \label{eq:Veff}
\end{align}

\section{Chameleon Behavior}

With a complete model in place, we now turn to a detailed investigation of the dynamics. We first examine the chameleon field, which is central to the effect we seek. Assuming that $\phi$ is homogeneous and isotropic, so that we can neglect spatial derivatives in $\Box\phi$, the equation of motion becomes
\begin{equation}
  \ddot{\phi} + 3H\dot{\phi} + V'(\phi) + \tilde{m}'_\psi(\phi) n_\psi = 0 \ .
  \label{eq:phi-eom-expanded}
\end{equation}
It is convenient for seeking numerical solutions to work with a dimensionless variable
\begin{equation}
  P \equiv \frac{\phi}{m_T}
\end{equation}
and to use $x$ as our independent variable. The equation of motion becomes
\begin{equation}
  P''(x) + \frac{2}{x} P'(x) + \frac{b^2 x^2}{m_T^3 g_*^\textrm{tot}}
  \left. \frac{dV}{d\phi} \right|_{\phi=Pm_T}
  + \frac{2\pi^2 b^2}{45x}\frac{g_{*S}}{g_*^\textrm{tot}}
  \left. \frac{d\tilde{m}_\psi}{d\phi} \right|_{\phi=Pm_T} Y(x) = 0 \ .
\end{equation}
We choose the initial conditions for $\phi$ to begin at the minimum of its effective potential and to move with same initial velocity as the changing minimum. The minimum $\phi_\textrm{min}$ solves the equation $V_\textrm{eff}'(\phi_\textrm{min})=0$, so one of the initial conditions for this equation can be obtained by evaluating this expression at $x_i$, using the relevant value for $Y(x_i)$ from~\eqref{eq:IC-Y}. Furthermore, since $\phi_\textrm{min}$ is a function of $x$, the initial velocity is found simply by taking a derivative and using the Boltzmann equation to obtain the relevant value for $Y'(x_i)$.

\subsection{Exponential Potentials}

Our goal here is to work out a single example model that exhibits the effects we are investigating, while at the same time remaining compatible with experimental constraints. For simplicity we will choose exponential functions, which also have the nice feature that observables approach a fixed asymptotic value at late times.

With these comments in mind, we therefore choose the form of the effective potential and U(1) coupling to be
\begin{subequations}
  \begin{align}
    V(\phi) &= \Lambda^4 e^{-\phi/m_1} \\
    \tilde{m}_\psi(\phi) &= m_\psi \left(1-A_2 e^{-\phi/m_2}\right) \\
    \tilde{f}(\phi) &= e\left(1+A_3 e^{-\phi/m_3}\right)^{-3} ,
  \end{align}
\end{subequations}
where $\Lambda$ and $m_\psi$ are constants with dimensions of mass, and $e$ and $A_2,\, A_3 > 0$ are dimensionless. The term with $A_2$ is necessary to incorporate the properties of $\psi$ into the equation of motion for $\phi$. The possibility for $A_3=0$ (constant gauge coupling) is viable, but we are specifically interested in increasing the cross section for $\psi$ as the universe expands. We choose this form for $\tilde{f}$ so that both the annihilation and scattering cross sections, which we calculate below, increase with time.

The largest energies of the particles in our theory are of order $m_\psi$ for nonrelativistic dark matter, since all other particles should be lighter than the dark matter to allow for annihilation. We, therefore, require $m_2,\, m_3 \gg m_\psi$ to suppress higher-dimensional operators involving derivatives of $\tilde{m}_\psi$ and $\tilde{f}$ when we expand the action. Additionally, we need $m_1 \gtrsim \Lambda$ to suppress higher-dimensional operators in the self-couplings of $\phi$.

The effective potential in \eqref{eq:Veff} is now
\begin{equation}
  V_\textrm{eff}(\phi)
  = \Lambda^4 e^{-\phi/m_1} + m_\psi (1-A_2 e^{-\phi/m_2})
  Y(x) \frac{2\pi^2}{45} g_{*S} \left(\frac{m_T}{x}\right)^3 \ ,
\end{equation}
possessing a critical point at
\begin{equation}
  \phi_\textrm{min}
  = -\frac{m_1 m_2}{m_2 - m_1} \ln\left(A_2 \frac{m_1}{m_2}
  \frac{m_\psi m_T^3}{\Lambda^4}\frac{Y}{x^3} \right) \ ,
  \label{eq:phi-min}
\end{equation}
which is real and finite. In order to generate a mass for the excitations of $\phi$, we require this critical point to be a minimum, which holds for
\begin{equation}
  m_2 > m_1 \ .
\end{equation}
The minimum moves with a speed
\begin{equation}
  \frac{d\phi_\textrm{min}}{dx} = -\frac{m_1 m_2}{m_2 - m_1}
  \left(Y \frac{dY}{dx} - \frac{3}{x}\right),
\end{equation}
which is positive ($\phi_\textrm{min}$ increases with $x$). Finally, we identify the initial conditions for $\phi$:
\begin{align}
  \phi(x_i) &= -\frac{m_1 m_2}{m_2 - m_1} \ln\left(A_2 \frac{m_1}{m_2}
  \frac{m_\psi m_T^3}{\Lambda^4}\frac{Y(x_i)}{x_i^3} \right) \ , \\
  \frac{d\phi}{dx}(x_i) &= -\frac{m_1 m_2}{m_2 - m_1}
  \left(Y(x_i) \frac{dY}{dx}(x_i) - \frac{3}{x_i}\right)  \nonumber \\
  &= \frac{m_1 m_2}{m_2 - m_1} \left(\frac{3}{x_i}
  + \frac{\tilde{m}_\psi}{\xi m_T}
  \frac{\Yeq(x_i) + (x_i^2 \tilde{m}_\psi)/(4B\xi m_T)}
       {\Yeq(x_i) + (x_i^2 \tilde{m}_\psi)/(2B\xi m_T)}
  \right) \ .
\end{align}
In order to ensure $\tilde{m}_\psi >0$, we require
\begin{equation}
  \phi > m_2 \ln(A_2)
\end{equation}
for all $\phi$ relevant for our calculation.

\subsection{An Attractor Solution}
A particularly interesting and simple possible evolution for the chameleon field is for it to begin at the minimum of the effective potential and then to adiabatically track this minimum as it evolves cosmologically. This attractor solution~\cite{Brax:2004qh} is achieved if the physical mass of the chameleon satisfies
\begin{equation}
  m_{\phi,\textrm{ph}} =
  \sqrt{V_\textrm{eff}^{\prime\prime}(\phi_\textrm{min})} \gg H \ .
  \label{eq:attractor-condition}
\end{equation}
If~\eqref{eq:attractor-condition} holds during radiation dominance, when
\begin{equation}
  H_R = \frac{m_T}{b} \sqrt{g_*^\textrm{tot}} x^{-2} \ ,
  \label{eq:H-rad}
\end{equation}
then we can avoid solving the coupled differential equations~\eqref{eq:Boltzmann-eq} and~\eqref{eq:phi-eom-expanded} and simply use the expression for $\phi_\textrm{min}$ for the evolution of $\phi$. Similarly, if~\eqref{eq:attractor-condition} holds during matter domination, when
\begin{equation}
  H_M = H_0 \left(\frac{x_0}{x}\right)^{3/2} \ ,
  \label{eq:H-matter}
\end{equation}
then we can easily determine $\phi_0$, the value of $\phi$ today, which is needed to calculate the values of the $\phi$-dependent parameters today.

Under the approximation that $m_2 \gg m_1$,
\begin{equation}
  \tilde{m}_{\phi,\textrm{ph}}
  \approx \left(A_2 \frac{2\pi^2}{45} \frac{m_\psi m_T^3}{m_1 m_2}\right)^{1/2}
  Y^{1/2} g_{*S}^{1/2} x^{-3/2} \ .
\end{equation}
It follows that $H_R$ decreases more rapidly than $\tilde{m}_{\phi,\textrm{ph}}$ with time, whereas during matter domination, $H_M$ and $\tilde{m}_{\phi,\textrm{ph}}$ have the same $x$ dependence. We shall verify later that these attractor solutions exist by numerically solving all the relevant equations of motion.

\section{Particle Physics Interactions and Constraints}
In the adiabatic regime described above, we now have all the ingredients necessary to understand the cosmological evolutions of the fields. We next turn to the particle physics phenomenology of the model. To do this, we rewrite the action~\eqref{eq:fullaction-dirac} without gravity to give the Lagrangian
\begin{align}
  \mathcal{L} \approx &
   - \frac{1}{2} \partial_\mu\phi \partial^\mu\phi - V(\phi)
  - (D_\mu\Phi)^\dagger (D^\mu\Phi) - V_0(\Phi) \nonumber  \\
  & {}-\frac{1}{4} F^{\mu\nu}F_{\mu\nu}
  + i\bar{\psi}\slashed{\partial}\psi - \tilde{m}_\psi(\phi) \bar{\psi}\psi
  - \tilde{f}(\phi) \bar{\psi} \slashed{A} \psi
  - \tilde{\lambda}_\psi(\phi) (\Phi + \Phi^\dagger) \bar{\psi} \psi \ ,
\end{align}
with $D_\mu = \partial_\mu + i\tilde{f}(\phi)A_\mu$.

\subsection{Breaking the Dark $U(1)$ Symmetry}
The potential of the dark Higgs field $\Phi$ is chosen so that this field acquires a vacuum expectation value (VEV)
\begin{equation}
  \left<0|\Phi(x)|0\right> = \frac{\vev}{\sqrt{2}} \ .
\end{equation}
Decomposing $\Phi$ into two real scalar fields via
\begin{equation}
  \Phi(x) = \frac{1}{\sqrt{2}}(\vev+h(x)) e^{-i\chi(x)/\vev} \ ,
\end{equation}
we can then use unitary gauge $\chi(x)=0$ to rewrite the kinetic term for $\Phi$ as
\begin{equation}
  -(D_\mu \Phi)^\dagger D^\mu \Phi = -\frac{1}{2} \partial^\mu h \partial_\mu h
  - \frac{1}{2} \tilde{f}^2(\phi) (\vev+h)^2 A^\mu A_\mu \ .
\end{equation}
Thus, the Goldstone boson is eaten to give the dark $U(1)$ gauge boson $A_\mu$ a mass $\tilde{M}_A(\phi)=\tilde{f}(\phi)\vev$. The Yukawa term generates a contribution to the mass of $\psi$, but since $\psi$ already has a Dirac mass, we need not rely on the dark Higgs to be the sole source of the $\psi$ mass. We, therefore, absorb the dark Higgs contribution into the definition of $\tilde{m}_\psi$ and retain the freedom to choose this mass scale and the coupling $\tilde{\lambda}_\psi(\phi)$ separately.

A typical choice for the pure dark Higgs potential $V_0(\phi)$ is
\begin{equation}
  V_0(\Phi) = \frac{1}{4} \tilde{\lambda}_h(\phi)
  \left[\Phi^\dagger \Phi - \frac{1}{2} \vev^2 \right]^2 \ ,
\end{equation}
which, when expanded about the VEV, yields
\begin{equation}
  V_0(h) = \frac{1}{4} \tilde{\lambda}_h(\phi) \vev^2 h^2
  + \frac{1}{4} \tilde{\lambda}_h(\phi) \vev h^3
  + \frac{1}{16} \tilde{\lambda}_h(\phi) h^4 \ .
\end{equation}
The mass of the physical dark Higgs particle $h$ is therefore
\begin{equation}
  \tilde{m}_h(\phi) = \sqrt{\frac{\tilde{\lambda}_h(\phi)}{2}}\vev \ ,
\end{equation}
and we see that the masses of the $A$ and $h$ fields are then related by
\begin{equation}
  \tilde{M}_A(\phi) = \tilde{f}(\phi)
  \sqrt{\frac{2}{\tilde{\lambda}_h(\phi)}} \tilde{m}_h(\phi) \ .
\end{equation}
Since the relative sizes of $\tilde{f}(\phi)$ and $\tilde{\lambda}_h(\phi)$ are unrestricted, in principle the relative masses of $A$ and $h$ are not fixed. However, in order to simplify the analysis, we will impose the hierarchy $\tilde{m}_h(\phi) > 2\tilde{M}_A(\phi)$ for all relevant $\phi$ so that $h$ has a tree-level decay channel to $A$.

Our Lagrangian at this stage is then
\begin{align}
  \mathcal{L} =
  & -\frac{1}{2} \partial_\mu\phi \partial^\mu \phi - V(\phi)
  - \frac{1}{2} \partial_\mu h \partial^\mu h
  - \frac{1}{4} \tilde{\lambda}_h(\phi) \vev^2 h^2
  - \frac{1}{4} \tilde{\lambda}_h(\phi) \vev h^3
  - \frac{1}{16} \tilde{\lambda}_h(\phi) h^4 \nonumber\\
  & {}- \frac{1}{4} F^{\mu\nu} F_{\mu\nu}
  -\frac{1}{2} \tilde{M}_A^2(\phi) A^\mu A_\mu
  + i\bar{\psi} \slashed{\partial} \psi
  - \tilde{m}_\psi(\phi) \bar{\psi} \psi
  - \tilde{f}(\phi) \bar{\psi}\gamma^\mu A_\mu \psi \nonumber\\
  & {} - \frac{1}{4} \left[2\tilde{f}^2(\phi)\right] h^2 A^\mu A_\mu
  - \frac{1}{2} \left[2\tilde{f}(\phi) \tilde{M}_A(\phi)\right] h A^\mu A_\mu
  - \sqrt{2} \tilde{\lambda}_\psi(\phi) h \bar{\psi}\psi \ .
  \label{eq:Ldirac}
\end{align}

What remains is to incorporate the fact that $\phi$ is adiabatically tracking the minimum of its effective potential. To achieve this, we expand $\phi(x)=\phi_c(t)+\eta(x)$ around its classical value and recall that $m_2$ and $m_3$ are sufficiently large to suppress non-relevant terms of $\order{\eta}$ or higher. The Lagrangian~\eqref{eq:Ldirac} then becomes
\begin{align}
  \mathcal{L} =
  & -\frac{1}{2} \partial_\mu\eta \partial^\mu \eta
  - \frac{1}{2} \partial_\mu h \partial^\mu h
  - \frac{1}{4} F^{\mu\nu} F_{\mu\nu}
  + i\bar{\psi} \slashed{\partial} \psi
  - \left[V(\phi_c) + \frac{1}{2}V''(\phi_c)\eta^2
    + \order{\eta^3} \right] \nonumber\\
  & {} - \frac{1}{4} \left[\tilde{\lambda}_h(\phi_c)
    + \order{\eta}\right] \vev^2 h^2
  - \frac{1}{6} \left[\frac{3}{2}\tilde{\lambda}_h(\phi_c)
    + \order{\eta}\right]  \vev h^3
  - \frac{1}{24} \left[\frac{3}{2}\tilde{\lambda}_h(\phi_c)
    + \order{\eta}\right] h^4 \nonumber\\
  & {} -\frac{1}{2} \left[\tilde{M}_A^2(\phi_c) +\order{\eta} \right] A^\mu A_\mu
  - \left[\tilde{f}(\phi_c) + \order{\eta}\right]
  \bar{\psi}\gamma^\mu A_\mu \psi \nonumber\\
  & {} - \left[\tilde{m}_\psi(\phi_c) + \tilde{m}'_\psi(\phi_c) \eta
    + \order{\eta^2}\right] \bar{\psi} \psi
  - \sqrt{2} \left[\tilde{\lambda}_\psi(\phi_c)
    + \order{\eta}\right] h \bar{\psi}\psi \nonumber\\
  & {} - \frac{1}{4} \left[2\tilde{f}^2(\phi_c)
    + \order{\eta}\right] h^2 A^\mu A_\mu
  - \frac{1}{2} \left[2\tilde{f}(\phi_c) \tilde{M}_A(\phi_c)
    + \order{\eta}\right] h A^\mu A_\mu  \ .
\end{align}

\subsection{The Dark Matter Annihilation Cross Section}

Our central goal is to understand how the dependence of dark matter cross sections on the chameleon field changes the standard dark matter creation, evolution, and detection story. To this end, we next turn to the calculation of the dark matter annihilation cross section. The relevant Feynman rules can be found in the Appendix.

\begin{figure}[t]
  \centering
  \includegraphics[scale=0.6]{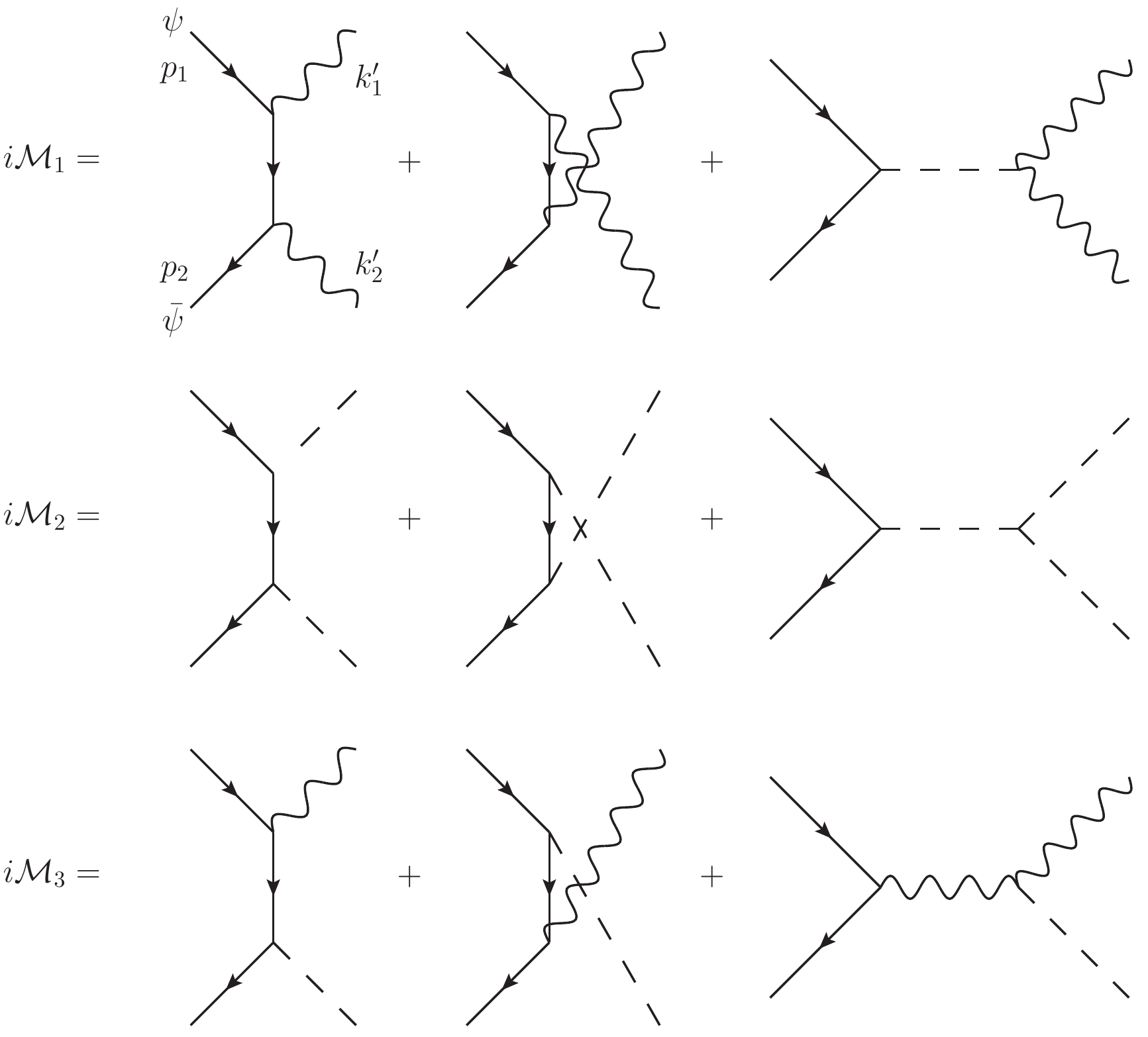}
  \caption{Tree-level $\psi$ annihilation diagrams. The massive vector boson $A$ is a wavy line, and the scalar $h$ is a dashed line. Annihilations to $A+A$ and $h+h$ via $\eta$-exchange and annihilations to final-state $\eta$ particles are suppressed by large-mass factors.}
  \label{fig:annihil-dirac}
\end{figure}

We assume that the dark matter is the heaviest particle in the dark sector, such that $\tilde{m}_\psi \gg \tilde{m}_h,~\tilde{M}_A$. Then, the lowest order, tree-level processes for $2\to 2$ dark matter annihilation are shown in Fig.~\ref{fig:annihil-dirac}, and their
amplitudes are
\begin{align}
  \mathcal{M}_1 = i \epsilon^\mu_{1'} \epsilon^\nu_{2'} \bar{v}_2
  \left[\tilde{f}^2(\phi_c) \left(\gamma_\nu \Delta_\psi(p_1-k'_1) \gamma_\mu
    \right.\right.
  &+ \left.\gamma_\mu \Delta_\psi(p_1-k'_2) \gamma_\nu \right) \nonumber \\
  &+ \left.\sqrt{2} \tilde{f}(\phi_c) \tilde{\lambda}_\psi(\phi_c)
    \tilde{M}_A(\phi_c) \Delta_h(p_1+p_2) g_{\mu\nu} \right] u_1
\end{align}
\begin{align}
  \mathcal{M}_2 = i \bar{v}_2 \left[2\tilde{\lambda}_\psi^2(\phi_c)
  \left(\Delta_\psi (p_1-k'_1) \right. \right.
  &+ \left. \Delta_\psi (p_1-k'_2)\right) \nonumber \\
  &+ \left. \frac{3}{\sqrt{2}} \tilde{\lambda}_\psi(\phi_c)
  \tilde{\lambda}_h(\phi_c) \vev \Delta_h(p_1+p_2) \right] u_1
\end{align}
\begin{align}
  \mathcal{M}_3 = i \epsilon_{1' \nu} \bar{v}_2
  \left[\sqrt{2}\tilde{f}(\phi_c)\tilde{\lambda}_\psi(\phi_c)
    \left(\Delta_\psi(p_1-k'_1)\gamma^\nu \right. \right.
  &+ \left. \gamma^\nu \Delta_\psi(p_1-k'_1)\right) \nonumber \\
  &+ \left. \tilde{f}^2(\phi_c)\tilde{M}_A(\phi_c) \gamma_\mu
    \Delta_A^{\mu\nu}(p_1+p_2) \right] u_1  \ .
\end{align}
Working in the center-of-mass frame and in the nonrelativistic limit we then obtain
\begin{align}
  \sigma_1 v &\approx
  \frac{\tilde{f}^4(\phi_c)}{16 \pi \tilde{m}_\psi^2(\phi_c)} \\
  \sigma_2 v &\approx
  \frac{15 \tilde{\lambda}_\psi^4(\phi_c)}{128\pi \tilde{m}_\psi^2(\phi_c)}v^2 \\
  \sigma_3 v &\approx
  \frac{\tilde{f}^2(\phi_c)\tilde{\lambda}_\psi^2(\phi_c)}
       {8\pi \tilde{M}_A^2(\phi_c)}
  =\frac{\tilde{\lambda}_\psi^2(\phi_c)}{8 \pi \vev^2} \ ,
\end{align}
where $v$ is the relative velocity of the incoming particles.

The interaction of main interest is between the dark matter and the gauge boson mediator. The dark Higgs's primary role is to break the U(1) symmetry to give the mediator a mass, and most of its particle interactions can be neglected. The contribution $\sigma_2$ is $p$ wave and thus subdominant to the other processes, which are $s$ wave. Also, the diagrams involving exchanges of $h$ in $\mathcal{M}_1$ and $\mathcal{M}_2$ do not significantly contribute. Thus, the dark Higgs has the opportunity to influence dark matter annihilations only via $\sigma_3$. Let us insist that the Yukawa coupling $\tilde{\lambda}_\psi$ is small enough (recall that the dark matter does not rely on this coupling to obtain a mass) such that $\sigma_3$ can be safely ignored. In this case only $\sigma_1$ remains and, since it is an $s$-wave cross section, it is a simple task to carry out the thermal averaging required in the Boltzmann equation. Note, however, that if thermal averaging is needed (following Ref.~\cite{Edsjo:1997bg}), we must use the dark sector temperature $T_d$ in the expression
\begin{equation}
  \left<\sigma v\right> = \frac{1}{(n_\psi^\textrm{EQ}(T_d))^2}
  \frac{g^2}{2(2\pi)^4} \int_{4\tilde{m}_\psi^2}^\infty ds\;
  \sqrt{s}T_d K_1\left(\frac{\sqrt{s}}{T_d}\right)
  (s-4\tilde{m}_\psi^2) \sigma(s) \ .
\end{equation}

\subsection{Corrections to the Cross Section}

\begin{figure}[t]
  \begin{tabular}{lcr}
    \includegraphics[width=0.3\textwidth]{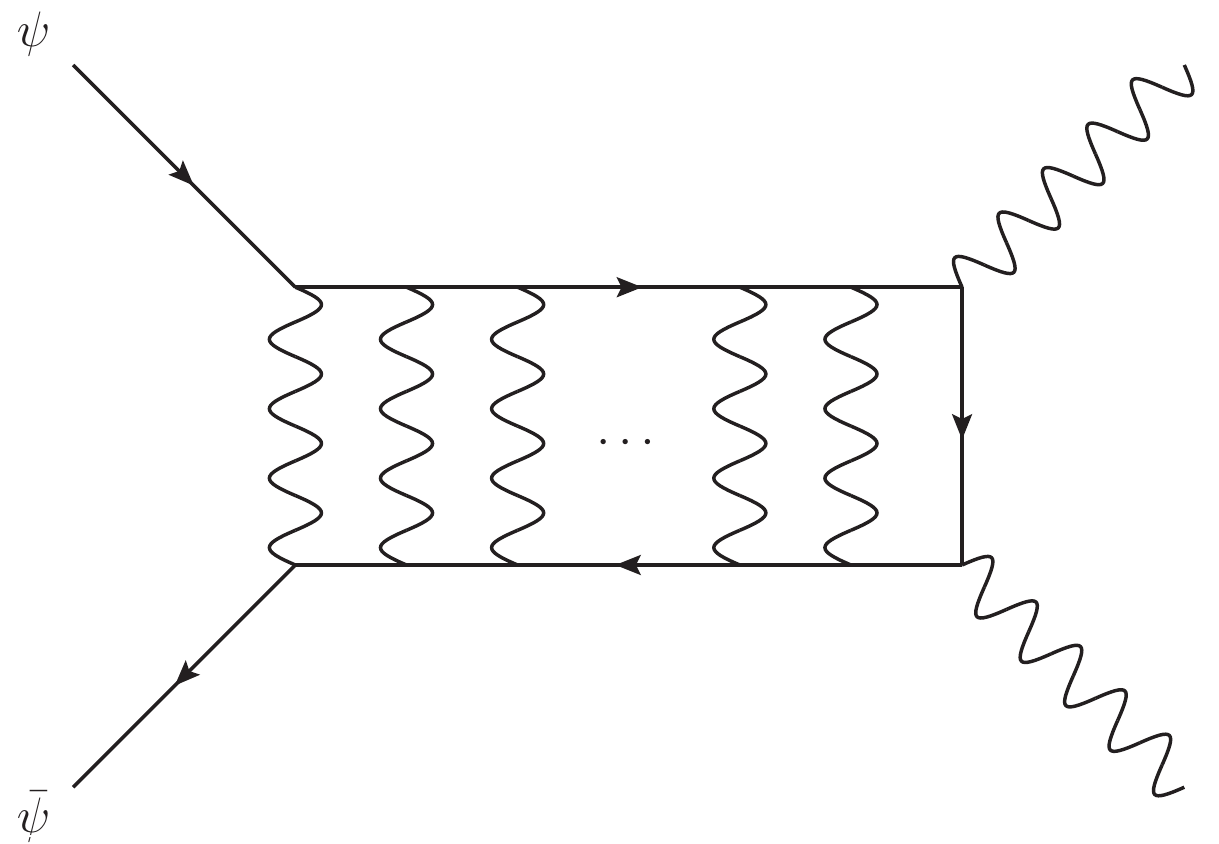} &
    \hspace{0.05\textwidth} &
    \includegraphics[width=0.3\textwidth]{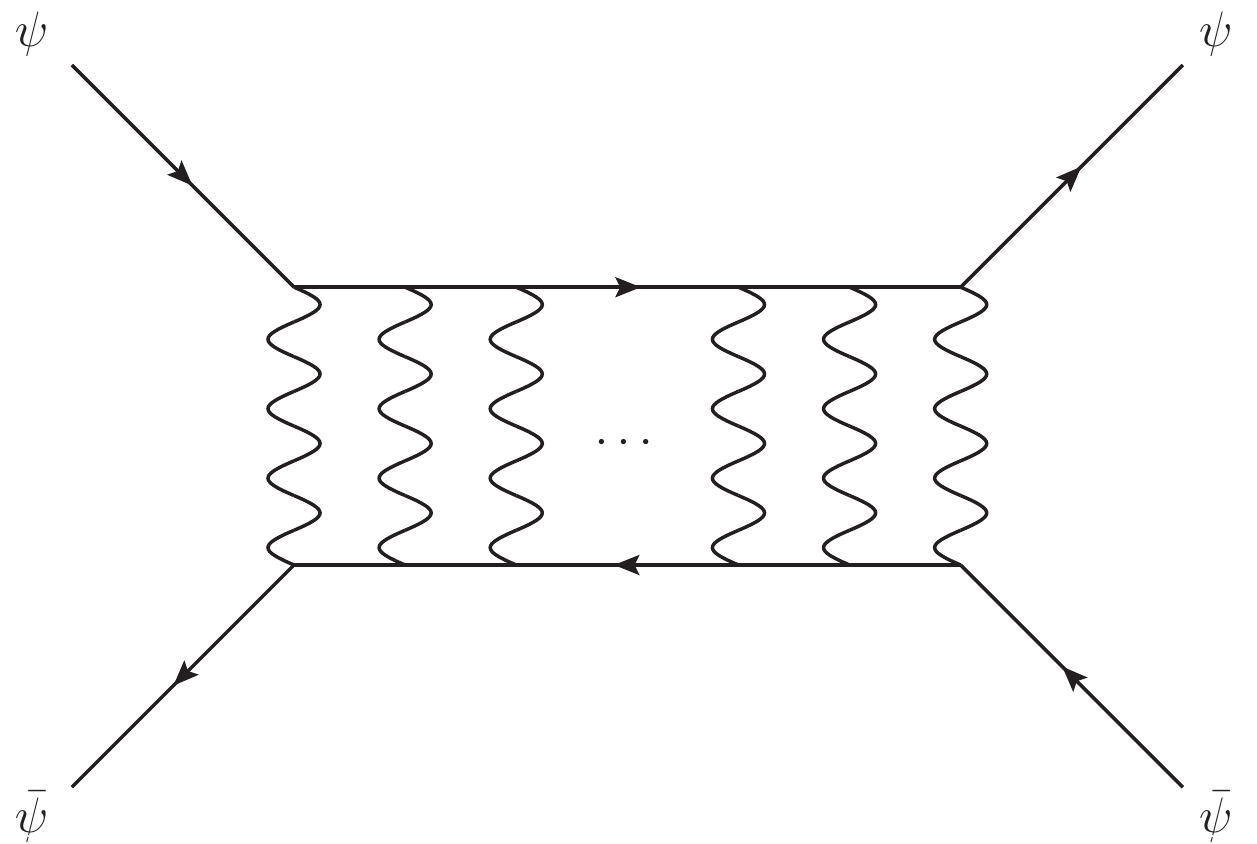}
  \end{tabular}
  \caption{Ladder diagrams for dark matter annihilation (left) and scattering (right).}
  \label{fig:ladder}
\end{figure}

We are interested in nonrelativistic dark matter, for which the relative velocities are much less than the speed of light. It is well known that for sufficiently low velocities, nonperturbative effects can have a large impact on the annihilation and scattering cross sections; and ladder diagrams, such as the ones shown in Fig.~\ref{fig:ladder}, must be included in the calculation.

\subsubsection{The Annihilation Cross Section}

In the case of annihilation, performing this summation is equivalent to solving the Schr\"{o}dinger equation in quantum mechanical scattering theory~\cite{Lifshitz-QED}.
This yields the so-called ``Sommerfeld enhancement''~\cite{Sommerfeld:1931} of the annihilation cross section (for detailed reviews in the context of dark matter, see, for example,~\cite{ArkaniHamed:2008qn,Lattanzi:2008qa,Robertson:2009bh}).
We consider the annihilation cross section $\sigma_0$ for a pointlike interaction near $r=0$ in perturbative field theory. For small velocities, the attractive Yukawa potential
\begin{equation}
  V(r) = -\frac{\tilde{\alpha}}{r} e^{-\tilde{M}_A r} \ ,
\end{equation}
where $\tilde{\alpha} = \tilde{f}^2(\phi_c)/4\pi$, distorts the wave function at the origin and cannot be ignored. Including the potential will enhance the annihilation cross section to $\sigma = \sigma_0 S_k$ by the Sommerfeld enhancement factor $S_k$. Let us define the dimensionless parameters
\begin{align}
  \epsilon_v &= \frac{v}{\tilde{\alpha}} \\
  \epsilon_A &= \frac{\tilde{M}_A}{\tilde{\alpha} \tilde{m}_\psi} \ ,
\end{align}
where $v$ is the velocity of each annihilating particle in the center of mass frame. In the case of a massless gauge boson with a Coulomb potential, it is possible to solve the Schr\"{o}dinger equation analytically to obtain the Sommerfeld enhancement.

For a massive gauge boson, the situation is more complicated, since the attractive potential has a finite range that limits the enhancement from being arbitrarily large for very low velocities. In the regime $\epsilon_A \ll \epsilon_v^2$, we recover the Coulomb case. At the crossover point $\epsilon_v \sim \epsilon_A$ (or equivalently $\tilde{m}_\psi v \sim \tilde{M}_A$), the de Broglie wavelength of the dark matter becomes comparable to the range of the interaction.  At lower velocities with $\epsilon_A \gg \epsilon_v^2$, the Yukawa potential cannot be ignored. As $v \to 0$, the de Broglie wavelength increases to a value larger than the interaction range, and thus the enhancement saturates at
\begin{equation}
  S_k \sim \frac{1}{\epsilon_A}
  \sim \frac{\tilde{\alpha} \tilde{M}_A}{\tilde{m}_\psi} \ .
\end{equation}
Furthermore, zero-energy bound states may form for certain values of $\epsilon_A$, giving resonance regions with larger enhancements $\sim \epsilon_A/\epsilon_v^2$ until they are cut off by finite width effects. In the early universe, freeze-out typically occurs at velocities $v_f \sim 0.3$, so that $\epsilon_v > 1$ and the Sommerfeld enhancement can be ignored. Note that there are no enhancements for $\epsilon_A > 1$.

To find the thermally averaged cross section, taking into account the Sommerfeld enhancement, we integrate $S_k$ using a Maxwellian distribution
\begin{equation}
  F(v) = \frac{4}{\bar{v}^3 \sqrt{\pi}} v^2 e^{-v^2/\bar{v}^2} \ ,
\end{equation}
where $\bar{v}$ is the characteristic velocity of the astrophysical system of
interest. Thus,
\begin{align}
  \left<\sigma v\right> &= (\sigma v)_{s\textrm{-wave}} \left<S_k\right> \\
  \left<S_k\right> &= \int_0^\infty dv\; F(v) S_k \ .
\end{align}

For the purposes of this paper, we choose to work in the $\epsilon_A > 1$ regime. This has two consequences. Practically, the calculation becomes much simpler, since we need not worry about the Sommerfeld enhancement at all. In addition, by deemphasizing the Sommerfeld enhancement, we clarify the extent to which the novel effects developed in this paper can alone increase the cross section over time in areas of parameter space that the Sommerfeld enhancement cannot reach.

\subsubsection{The Scattering Cross Section}
To find the scattering cross section, we can use nonrelativistic quantum mechanics and sum over partial waves. The total cross section is
\begin{equation}
  \sigma = \frac{4\pi}{k^2} \sum_{l=0}^{\infty} (2l+1) \sin^2\delta_l \ ,
\end{equation}
although a more useful quantity to compare to observational constraints is the transfer cross section
\begin{align}
  \sigma_\mathrm{tr} &= \int d\Omega\; (1-\cos\theta) \frac{d\sigma}{d\Omega} \nonumber  \\
  &= \frac{4\pi}{k^2} \sum_l [(2l+1)\sin^2\delta_l
    - 2(l+1)\sin\delta_l \sin\delta_{l+1}\cos(\delta_{l+1}-\delta_l)] \ ,
\end{align}
which controls the rate at which energy is transferred between colliding particles. Following~\cite{Buckley:2009in}, analytic estimates for the cross section are
\begin{align}
  \sigma &= \frac{4\pi}{\mu^2 v_\textrm{rel}^2} (1+L)^2 \\
  \sigma_\mathrm{tr} &= \frac{4\pi}{\mu^2 v_\textrm{rel}^2} (1+L) \ ,
\end{align}
where $L=\mu v_\textrm{rel} b_\textrm{max}$ is the largest angular momentum needed to describe the interaction between two particles of reduced mass $\mu=\tilde{m}_\psi/2$ that travel with a relative velocity $v_\textrm{rel}$ and maximum relevant impact parameter $b_\textrm{max}$. Note that these estimates are only valid for $L \gtrsim 1$. We estimate the impact parameter by solving
\begin{equation}
  \frac{1}{2} \mu v_\textrm{rel}^2
  = \frac{\tilde{f}^2/4\pi}{b_\textrm{max}} e^{-\tilde{M}_A b_\textrm{max}}  \ .
\end{equation}

If we work in the $\epsilon_A > 1$ regime to avoid Sommerfeld enhancements, then we will also tend to avoid enhancements to the scattering cross section and can expect to be working in the Born limit. Simply taking the nonrelativistic limit of the perturbative cross section gives
\begin{equation}
  \sigma = \frac{\tilde{f}^4(\phi_c) \tilde{m}_\psi^2(\phi_c)}
         {8\pi \tilde{M}_A^4(\phi_c)}
         = \frac{\tilde{m}_\psi^2(\phi_c)}{8\pi \vev^4}\ .
\end{equation}
Assuming that dark matter self-interactions are not needed to explain the structure of dwarf galaxies~\cite{Buckley:2009in}, we use a conservative bound~\cite{Hannestad:2000bs} (see also \cite{Rocha:2012,Peter:2012})
\begin{equation}
  \sigma/\tilde{m}_\psi < 0.1 ~\units{cm^2/g}
  \label{eq:scatterlimit}
\end{equation}
for characteristic velocities of $10~\units{km/s}$. As we mention below, it would not be difficult to find parameters that violate this bound.

In the usual treatment of dark matter, constraints such as this one, obtained from present-day observations, can be directly applied to bounds on physics at freeze-out or before. It is important to remember here that, in our model, the evolution of the chameleon field means that such a connection is far less direct, and such bounds typically do not apply in the early universe.

\subsection{Dark Decays}
\begin{figure}[t]
  \centering
  \includegraphics[scale=0.6]{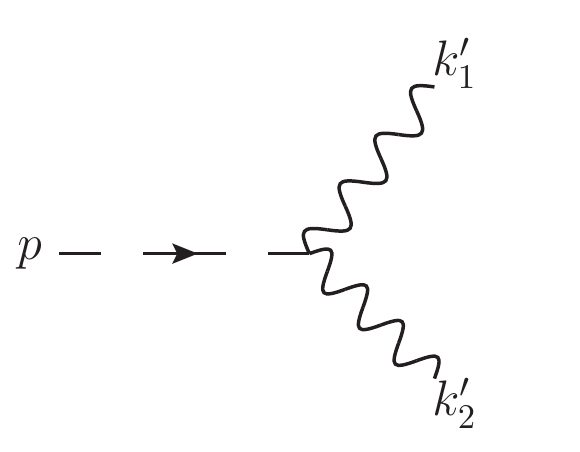}
  \caption{Tree-level $h$ decay.}
  \label{fig:hdecay-dirac}
\end{figure}

The dark Higgs $h$ and the dark gauge boson $A$ are allowed to decay. As mentioned earlier, we assume $\tilde{m}_h(\phi_c) > 2 \tilde{M}_A(\phi_c)$ so that $h$ has a tree-level decay channel to $A$, as shown in Fig.~\ref{fig:hdecay-dirac}.
Its decay width is then
\begin{equation}
  \Gamma_h = \frac{\tilde{f}^2(\phi_c)}{32\pi}
  \frac{\tilde{m}_h^3(\phi_c)}{\tilde{M}_A^2(\phi_c)}
  \sqrt{1-\frac{4\tilde{M}_A^2(\phi_c)}{\tilde{m}_h^2(\phi_c)}}
  \left[1-\frac{4\tilde{M}_A^2(\phi_c)}{\tilde{m}_h^2(\phi_c)}
    + 12\frac{\tilde{M}_A^4(\phi_c)}{\tilde{m}_h^4(\phi_c)}\right] \ .
\end{equation}

\begin{figure}[t]
  \centering
  \includegraphics[scale=0.6]{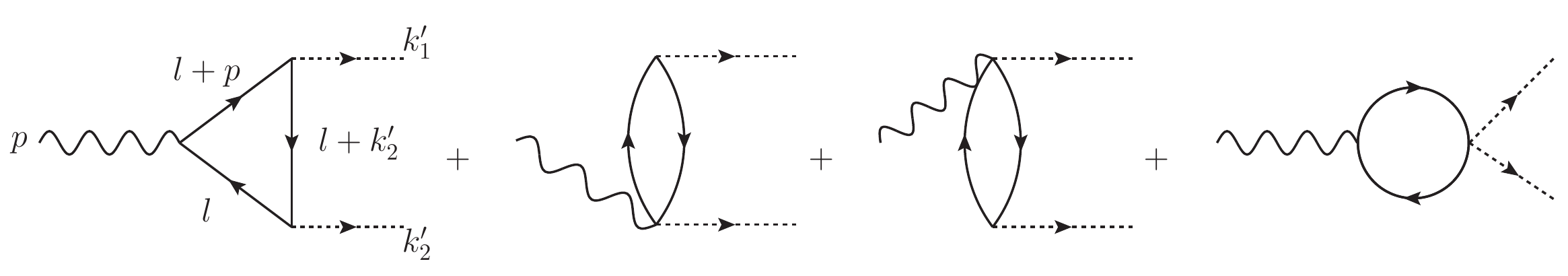}
  \caption{1-loop $A$ decay. Only the first diagram is nonzero.}
  \label{fig:Adecay-dirac}
\end{figure}

Although the $A$ particle is allowed to decay to $\eta$ particles, which are substantially smaller in mass, this occurs through a 1-fermion-loop process, as shown in Fig.~\ref{fig:Adecay-dirac}. The amplitude is also suppressed by two factors of $m_2$ from the $\order{\eta^2}$ term in the expansion of $\tilde{m}_\psi(\phi)$.
The nonzero amplitude in the limit of $\tilde{m}_\eta \ll \tilde{M}_A,\, \tilde{m}_\psi$ is
\begin{align}
  \mathcal{M} =
  -\frac{4i\pi^2 \tilde{f} \left(\tilde{m}^\prime_\psi\right)^2}{\tilde{M}_A^2}
  k'_2 \cdot \epsilon^*(p) & \left[4\tilde{m}_\psi \tilde{M}_A
  C_0[p^2,(p-k'_2)^2,k_2^2,\tilde{m}_\psi,\tilde{m}_\psi,\tilde{m}_\psi]
  \right. \nonumber  \\
  & \left. + (8\tilde{m}_\psi^2 + \tilde{M}_A^2)
  B_0[p^2,\tilde{m}_\psi,\tilde{m}_\psi]
  -8\tilde{m}_\psi^2 B_0[k_2^{\prime 2},\tilde{m}_\psi,\tilde{m}_\psi] \right] \ ,
  \label{eq:Adecay-amp}
\end{align}
where $B_0$ and $C_0$ are scalar Passarino-Veltman functions ~\cite{tHooft:1978xw,Passarino:1978jh,Consoli:1979xw,Denner:1991kt}, defined via
\begin{align}
  B_0[p^2,m^2,m^2] &=
  \frac{1}{i\pi^2} \int d^4l\; \frac{1}{(l^2+m^2)[(l+p)^2+m^2]} \\
  C_0[p^2,(p-p_1)^2,p_1^2,m,m,m] &= -\frac{1}{i\pi^2} \int d^4l\;
  \frac{1}{(l^2+m^2)[(l+p)^2+m^2][(l+p_1)^2+m^2]} \ .
\end{align}
The $C_0$ integral is finite and, in the approximation $\tilde{m}_\psi \gg \tilde{M}_A \gg \tilde{m}_\eta$, reduces to
\begin{equation}
  C_0[p^2,(p-k'_2)^2,k_2^{\prime 2},\tilde{m}_\psi,\tilde{m}_\psi,\tilde{m}_\psi]
   \approx -\frac{1}{4\tilde{m}_\psi^2} \ .
\end{equation}
The $B_0$ integral diverges, so we cut off the loop-momentum integral at some large scale. Using $m_3$ for this purpose, since we will often find it numerically to be the largest mass-suppression scale in our theory, we have
\begin{equation}
  B_0[p^2,\tilde{m}_\psi,\tilde{m}_\psi]
  \approx B_0[k_2^{\prime 2},\tilde{m}_\psi,\tilde{m}_\psi]
  \approx 2\ln\left(\frac{m_3}{\tilde{m}_\psi}\right) \ .
\end{equation}
Putting everything together, the decay width of $A$ is then given by
\begin{equation}
  \Gamma_A \approx \frac{1}{6\tilde{M}_A} \pi^2 \tilde{f}^2
  \left(\frac{m_\psi A_2}{m_2}\right)^4 e^{-4\phi_c/m_2}
  \left[\ln\left(\frac{m_3}{\tilde{m}_\psi}\right)\right]^2 \ .
\end{equation}
The $A$ bosons must decay efficiently enough not to contribute significantly to the energy density budget today. Though the decaying exponential makes meeting this criterion more difficult, there is still a small sample of parameter space for which the $A$ energy density does not pose a problem.

\section{Numerical Solutions}

While we have described a number of ways to understand the evolution of the fields analytically, including, for example, the adiabatic approximation in which the chameleon tracks the minimum of its effective potential, ultimately, we are able to numerically solve the relevant equations of motion completely. To do so, of course, we must make sensible choices for our parameters to satisfy the various bounds and simplifying inequalities we have specified.

We need to implement the correct relationship between the dark sector temperature $T_d$ and that in the photon sector $T$, which in turn requires us to correctly enumerate the massless degrees of freedom at the relevant scales. At the unification scale, all the dark particles ($\psi$, $A$, $h$, $\phi$) are relativistic, so $g_{*S}^d(t_u)=8.5$. Around the epoch of dark matter freeze-out, only $\psi$ is nonrelativistic, so $g_{*S}^d(t_f)=5$. Thus, at freeze-out, $\xi_f = 1.19$ for $g_{*S}(t_f)=106.75$ or $\xi_f = 0.56$ for $g_{*S}(t_f)=10.75$. With these numbers, the bound on the number of effective neutrino species in \eqref{eq:neutrino-bounds} is easily satisfied.

The model is insensitive to $\tilde{M}_A(\phi)$ and $\tilde{m}_h(\phi)$ at lowest order. We choose $\vev$ such that $\bar{\psi}\psi \to AA$ is kinematically allowed today, while ensuring $\epsilon_A >1$ and $\tilde{\alpha}<1$. We must then check that $\tilde{M}_A(\phi_0)$ satisfies scattering cross-section bounds. It is simplest to assume the attractor solution for $\phi$ and then later verify that it is in fact adhered to. The $A$ gauge bosons need to decay away before BBN so that their energy density is negligible. Finally, we must ensure that the evolution ends with the observed density of dark matter today. For this figure we use the bounds from the 7-year WMAP data~\cite{Larson:2010gs}, assuming a $\Lambda$CDM cosmology
\begin{equation}
  \Omega_\textrm{DM} h^2 = \frac{\rho_0}{\rho_{c0}} = 0.1109 \pm 0.0056 \ .
\end{equation}

\begin{figure}[t]
  \centering
  \includegraphics[width=0.98\textwidth]{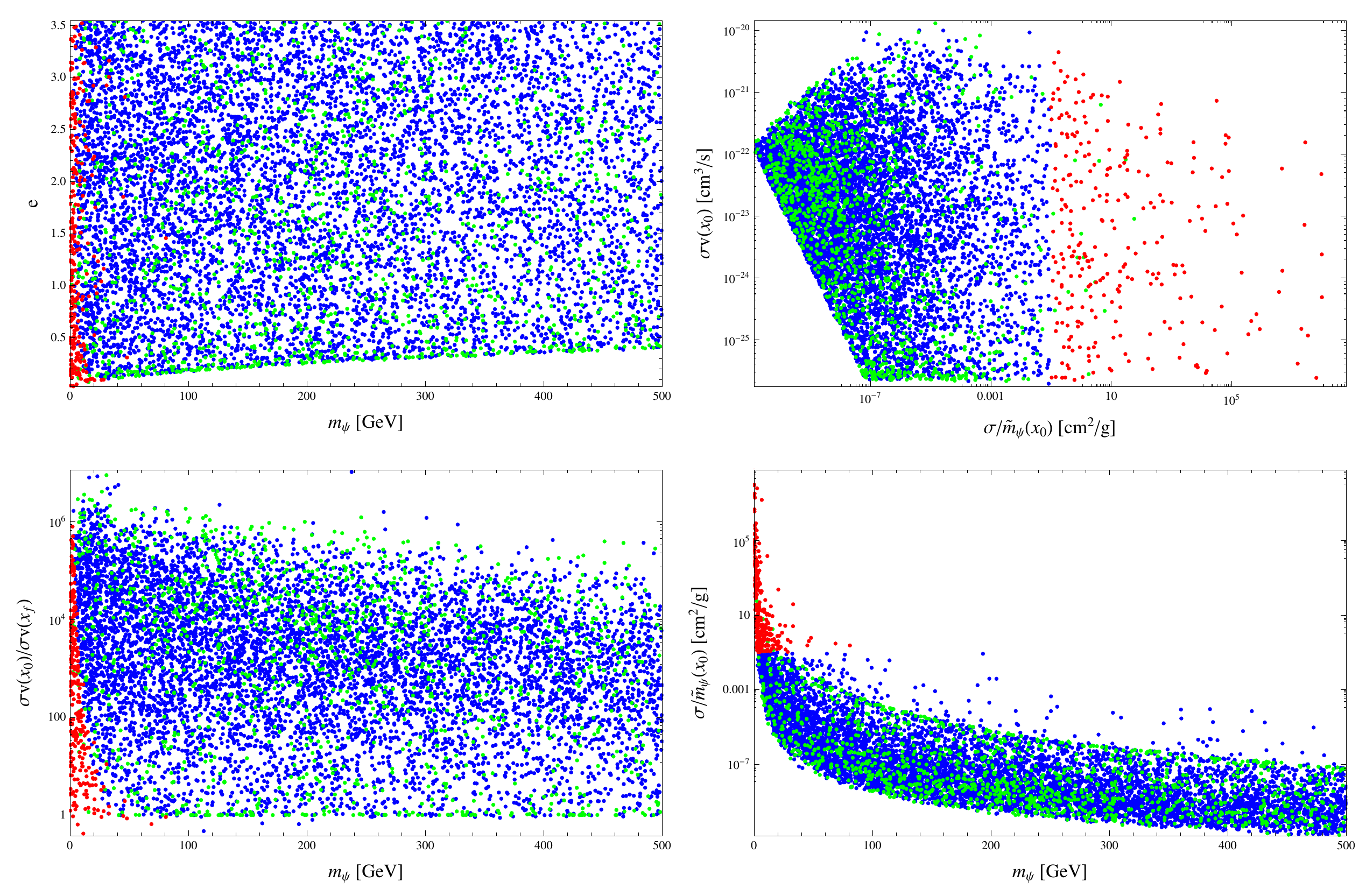}
  \caption{Scan of parameter space. Blue points indicate sets of parameters that satisfy all constraints, except (for most points) for having a negligible $A$ energy density. Red points do not satisfy the scattering cross section bound $\sigma/\tilde{m}_\psi < 0.1~\units{cm^2/g}$. Green points do not satisfy the adiabatic condition in \eqref{eq:attractor-condition} and should be solved with the coupled differential equations.}
  \label{fig:paramscan}
\end{figure}
\begin{figure}[t]
  \centering
  \includegraphics[scale=0.6]{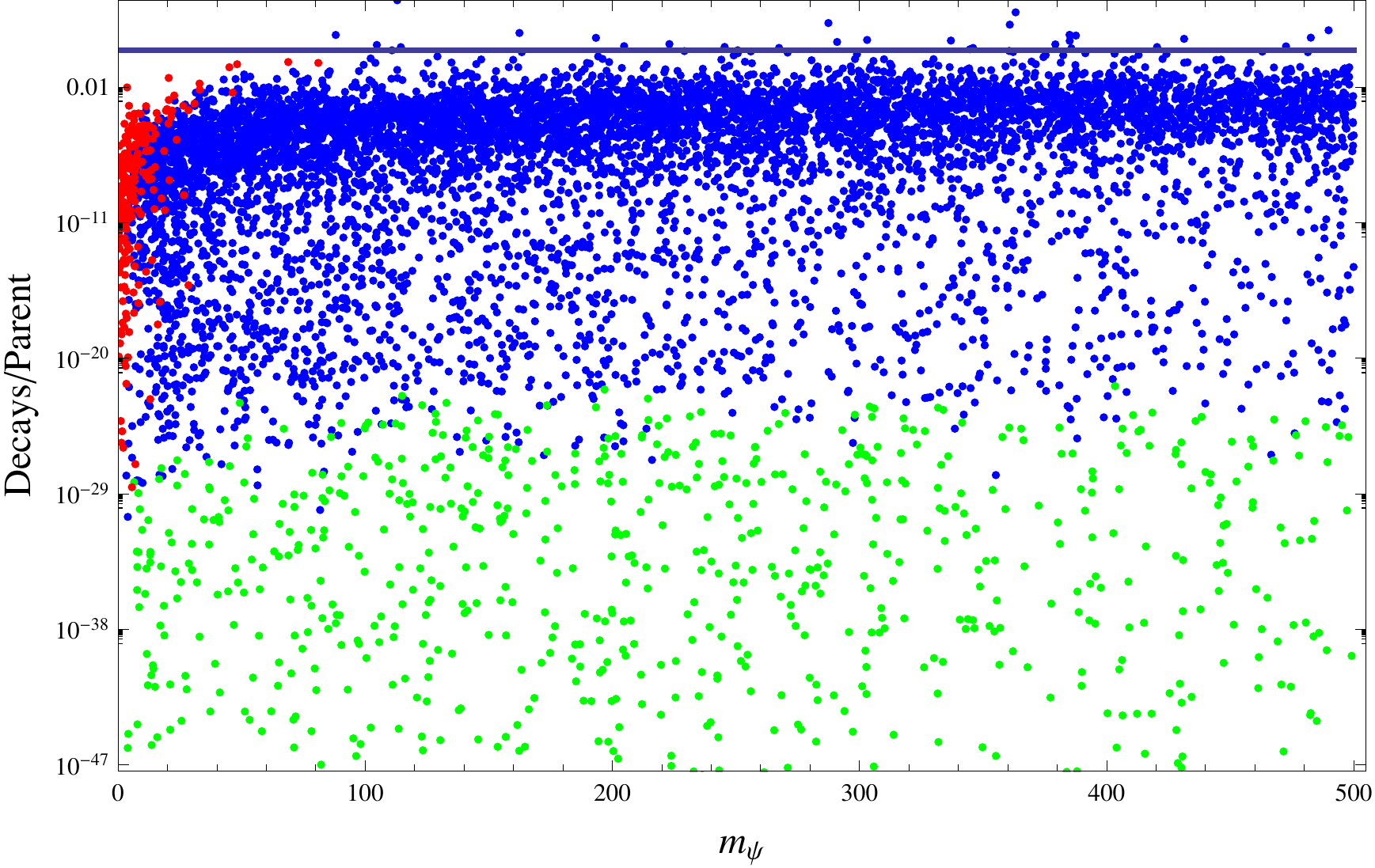}
  \caption{The number of $A$ decays per particle between freeze-out and BBN. Points above the horizontal line at $1$ indicate that all $A$ particles should have decayed and thus do not contribute significantly to the energy budget of the Universe.}
  \label{fig:Adecays}
\end{figure}

Given these constraints, we numerically solve the Boltzmann equation and show a sample of parameter space in Fig.~\ref{fig:paramscan}, resulting from a random, uniform scan over $m_\psi \in [0.1,500]~\units{GeV}$; $m_1 \in [10^5, 10^7]~\units{GeV}$; $m_2 \in [5\times 10^5, 5\times 10^8]~\units{GeV}$; $m_3 \in [5\times 10^5, 5\times 10^8]~\units{GeV}$; $\Lambda \in [10, 10^3]~\units{GeV}$; $A_2 \in [0.1, 9.9]$; $A_3 \in [0.1, 10]$; and $e \in [0.01, \sqrt{4\pi}]$. The upper-left panel shows the coupling parameter $e$ vs the dark matter mass parameter $m_\psi$. The upper-right panel shows the annihilation cross section $\sigma v$ vs the scattering cross section $\sigma/\tilde{m}_\psi$, both evaluated at $x_0$ today. The bottom panels show the boost in annihilation cross section from freeze-out to today and the scattering cross section today vs the mass parameter $m_\psi$. Again, there is flexibility when choosing $\vev$ without affecting the evolution of $\phi$ and $Y$ at lowest order, so it is possible to obtain valid models for a scaled value of $\sigma/\tilde{m}_\psi$. Here, we show the largest possible scattering cross sections, while staying within the bound $\epsilon_A >1$. As demonstrated in Fig.~\ref{fig:Adecays}, only a small portion of the sampled parameter space fulfills the requirement that the $A$ gauge boson energy density is negligible by the time of BBN. While finding a set of parameters that satisfies all constraints is certainly possible, the effect of having very large increases in the annihilation cross section does not seem to be a general feature of the model.

\begin{figure}[t]
  \centering
  \includegraphics[width=0.98\textwidth]{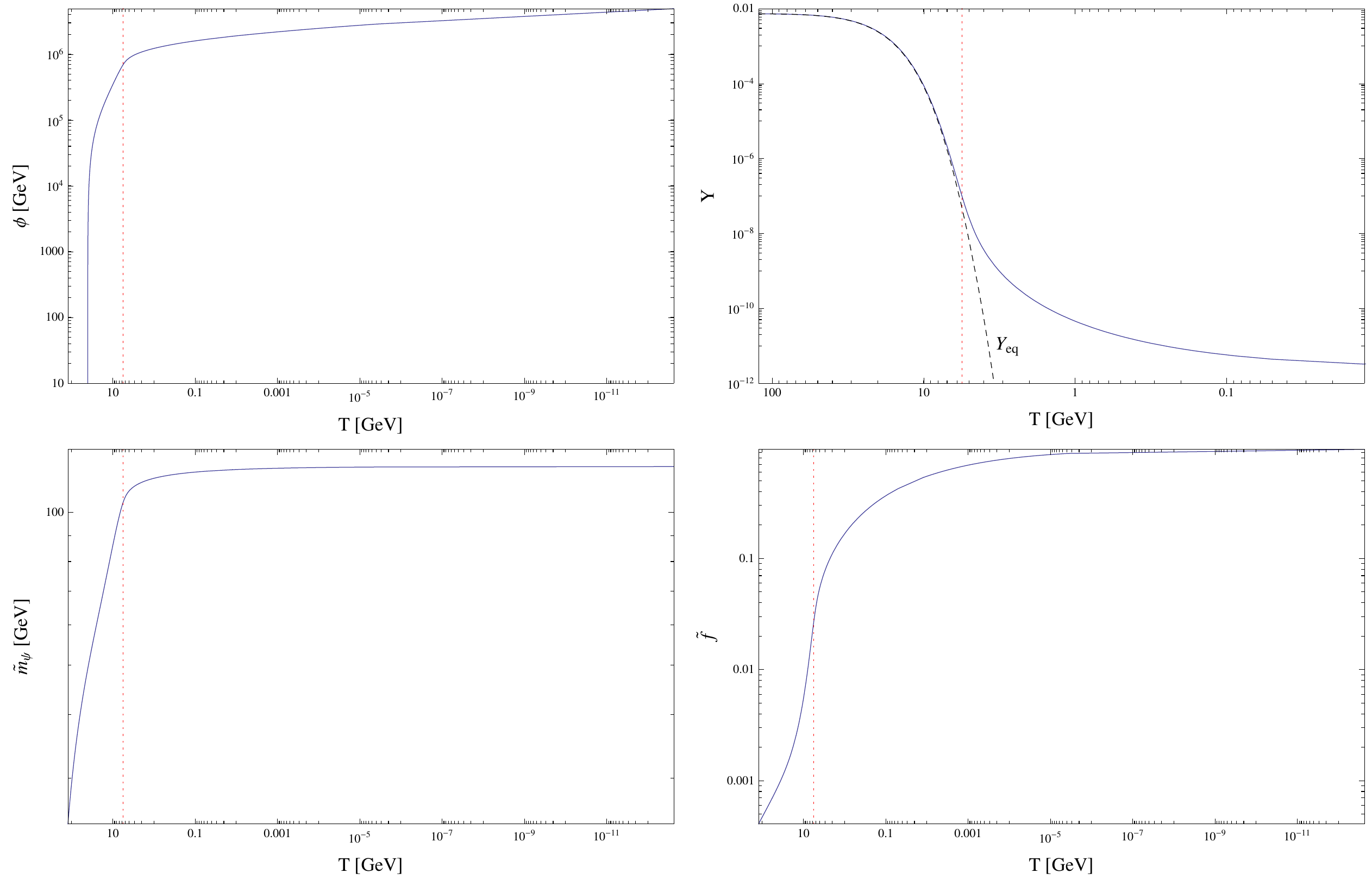}
  \caption{The evolution of $\phi$ (upper left), $Y$ (upper right), $\tilde{m}_\psi$ (lower left), and $\tilde{f}$ (lower right) as a function of $T$ in GeV for the model with $m_\psi=123$~GeV. The red, dotted line indicates the approximate dark matter freeze-out temperature.}
  \label{fig:param-evolution}
\end{figure}
\begin{figure}[t]
  \centering
  \includegraphics[width=0.98\textwidth]{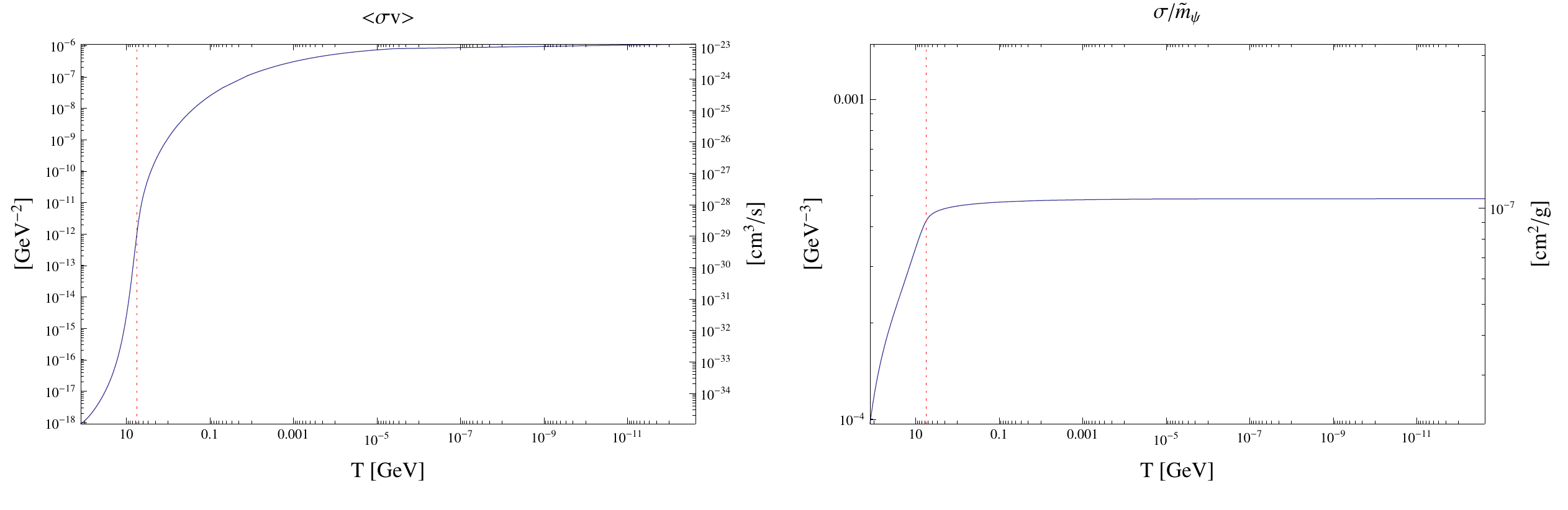}
  \caption{The evolution of the annihilation cross section $\langle\sigma v\rangle$ (left) and the scattering cross section $\sigma/\tilde{m}_\psi$ (right) as a function of $T$ in GeV with $m_\psi=123$~GeV. The red, dotted line indicates the approximate dark matter freeze-out temperature.}
  \label{fig:crossSec-evolution}
\end{figure}
As a concrete example, we show a specific model with the parameter choices: $m_{\psi}=123$~GeV; $m_T=m_{\psi}$; $m_1=38$~TeV; $m_2=500$~TeV; $m_3=500$~TeV; $\Lambda=18$~GeV; $A_2=0.6$; $A_3=9.2$; $e=0.96$; and $\vev=10$~GeV. This comprises an optimistic set of parameter choices, that satisfies all our bounds and provides a large change of order $\sim 10^6$ in the annihilation cross section over the history of the Universe. Our choice for the value of $\vev$ gives $\epsilon_A=1.07$ today, and we can ignore the Sommerfeld enhancement. Larger values of $\vev$ work equally well; increasing $\vev$ increases $\epsilon_A \sim \vev$ and decreases $\sigma/\tilde{m}_\psi \sim \vev^{-4}$. The dark matter relic density is $\Omega_\psi h^2 = 0.1097$, within a standard deviation of the observed value. The scattering cross section today is $4.9\times 10^{-4}~\units{cm^2/g}$, well below the conservative limit in \eqref{eq:scatterlimit}. We must also check that these parameters satisfy the assumptions we have made in writing down the model. For example, we neglected terms with $\partial_\mu \tilde{f}/\tilde{f}$, and here we note that $\dot{\tilde{f}}/\tilde{f} \sim 10^{-9} - 10^{-6}~\units{GeV}$, which is much smaller than other mass terms in the perturbative expansion. The adiabatic approximation is satisfied with $H/m_{\phi,\textrm{ph}} \sim 10^{-11}$ throughout the evolution of $\phi$. Finally, we use the decay width of the $A$ particles to determine that they have decayed away in the time from freeze-out to BBN, so they do not contribute to the energy budget we observe from the CMB.

The results for the evolution of $\phi$, $Y$, the dark matter mass $\tilde{m}_\psi$, and the coupling $\tilde{f}$ as a function of $T=m_T/x$ are shown in Fig.~\ref{fig:param-evolution}. We also show the annihilation and scattering cross sections in Fig.~\ref{fig:crossSec-evolution}. The scattering cross section quickly approaches its asymptotic value by the time of dark matter freeze-out, while the annihilation cross section still grows orders of magnitude from freeze-out to now. This difference is due to the scattering cross section, $\sigma/\tilde{m}_\psi \propto \tilde{m}_\psi/\vev^4$, and the annihilation cross section, $\sigma v \propto \tilde{f}^4/\tilde{m}_\psi^2$, depending differently on $\phi$ via $\tilde{m}_\psi$ and $\tilde{f}$. We choose the form of $\tilde{f}$ to force the annihilation cross section to grow more slowly, whereas the scattering cross section has no such term countering its growth. With these particular choice of parameters, the scattering cross section is too small to have interesting astrophysically observable consequences.

As shown in Fig.~\ref{fig:paramscan}, there are other choices of parameters that will still give a boost to the annihilation cross section while yielding a larger scattering cross section to match observational bounds \cite{Rocha:2012,Peter:2012}; however, again, most of the plotted parameter space is restricted from the $A$ energy density requirement. One option for increasing the viable parameter space is to relax the requirement that $\epsilon_A >1$ and to work in the regime of Sommerfeld enhancements; our model would still provide significant increases in the cross sections, and Sommerfeld enhancements would serve to further increase the boosts. Another clear option is to open an alternative decay channel for $A$.

\section{Conclusions}
In this paper we have investigated the possibility that the properties of dark matter depend crucially on the dynamics of a chameleon field -- a scalar field whose cosmological evolution depends not only on its bare potential but also on the local density of other matter (such as dark matter itself) in the Universe. We have shown that such a coupling allows the annihilation cross section (for example) of the dark matter particles to change by several orders of magnitude between freeze-out and today, while remaining consistent with all observational constraints. We have presented a general formalism to describe how this might happen and have provided a specific particle physics example, in which all relevant quantities can be calculated. While there are significant observational and theoretical constraints on models of this type, it is nevertheless possible for the cross section to evolve in such a way that there may be interesting implications for the detection of dark matter and for its dynamical effects on late-universe astrophysics.

There are, of course, other possible complications to this idea that are beyond the scope of the current paper but that provide interesting avenues for future study. One natural step is to couple our model directly to the Standard Model. One way to achieve this is to directly add the dark U(1) to the current SM gauge group~\cite{Gondolo:2011eq}. Another possibility is to couple to the Standard Model through $U(1)$ kinetic mixing ~\cite{Holdom:1985ag,Feng:2010zp}. This extension of our model should be able to easily accommodate the relevant particle physics constraints ~\cite{Bjorken:2009mm,Batell:2009yf,Pospelov:2008jd,Pospelov:2008zw}, while easily allowing for decays of the dark gauge boson to Standard Model particles well before BBN. The dark matter annihilations would still be dominated by the channel $\bar{\psi}\psi \to AA$, since annihilation to Standard Model particles would be suppressed by the small coupling parameter for the U(1) mixing. However, it is a more delicate issue to decide what a natural route would be to couple the visible and dark scalar sectors, particularly with regards to coupling the chameleon to normal matter.

Finally, we did not attempt a careful analysis of the effect of late-universe inhomogeneities on the chameleon field or the dark matter properties on which it depends. In the specific models we considered, it seems as if such effects would be small, but a more careful examination is warranted.

\begin{acknowledgments}
We use Mathematica 8 to numerically solve the Boltzmann equation, JaxoDraw 2.1-0~\cite{JaxoDraw-v2,JaxoDraw-v1} to create the Feynman diagrams and FeynCalc~\cite{FeynCalc} to calculate the 1-loop $A$ decays. We would like to acknowledge helpful discussions with Matt Buckley, Katie Freese, Koji Ishiwata, Justin Khoury, and Mark Wise. This work was supported in part by the Department of Energy, the National Science Foundation, NASA, and the Gordon and Betty Moore Foundation.

\end{acknowledgments}

\newpage
\appendix*
\section{Feynman Rules}
\label{sec:FeynRules-dirac}
\begin{figure}[h]
  \centering
  \includegraphics[scale=0.8]{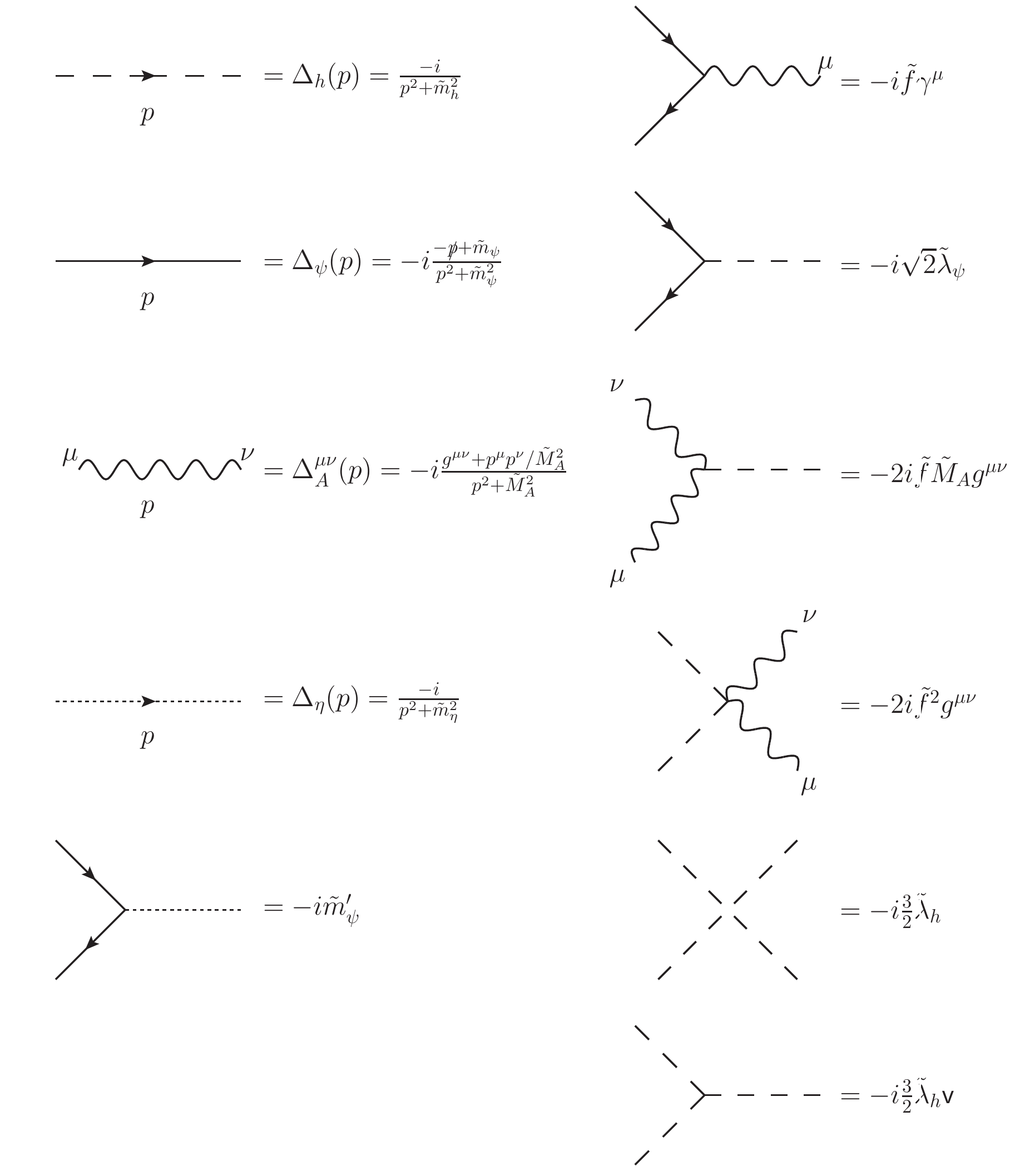}
  \caption{Feynman rules for $h$ (dashed line), $\psi$ (solid line), $A_\mu$
    (wavy line), and $\eta$ (dotted line). We include the Yukawa interaction with $\psi$ and $\eta$, which is relevant for the 1-loop $A$-decay amplitude in \eqref{eq:Adecay-amp}, but other $\eta$-interaction vertices are not shown. All parameters labeled by a tilde are evaluated at $\phi_c$.}
  \label{fig:Frules-dirac}
\end{figure}

The Feynman rules are shown in Fig.~\ref{fig:Frules-dirac}. All of these diagrams have higher-order corrections that involve $\eta$ particles.


\vfill\eject
\bibliography{bibliography}

\begin{thebibliography}{56}
\expandafter\ifx\csname natexlab\endcsname\relax\def\natexlab#1{#1}\fi
\expandafter\ifx\csname bibnamefont\endcsname\relax
  \def\bibnamefont#1{#1}\fi
\expandafter\ifx\csname bibfnamefont\endcsname\relax
  \def\bibfnamefont#1{#1}\fi
\expandafter\ifx\csname citenamefont\endcsname\relax
  \def\citenamefont#1{#1}\fi
\expandafter\ifx\csname url\endcsname\relax
  \def\url#1{\texttt{#1}}\fi
\expandafter\ifx\csname urlprefix\endcsname\relax\def\urlprefix{URL }\fi
\providecommand{\bibinfo}[2]{#2}
\providecommand{\eprint}[2][]{\url{#2}}

\bibitem[{\citenamefont{Carlson et~al.}(1992)\citenamefont{Carlson, Machacek,
  and Hall}}]{Carlson:1992fn}
\bibinfo{author}{\bibfnamefont{E.~D.} \bibnamefont{Carlson}},
  \bibinfo{author}{\bibfnamefont{M.~E.} \bibnamefont{Machacek}},
  \bibnamefont{and} \bibinfo{author}{\bibfnamefont{L.~J.} \bibnamefont{Hall}}
  (\bibinfo{year}{1992}).

\bibitem[{\citenamefont{de~Laix et~al.}(1995)\citenamefont{de~Laix, Scherrer,
  and Schaefer}}]{deLaix:1995vi}
\bibinfo{author}{\bibfnamefont{A.~A.} \bibnamefont{de~Laix}},
  \bibinfo{author}{\bibfnamefont{R.~J.} \bibnamefont{Scherrer}},
  \bibnamefont{and} \bibinfo{author}{\bibfnamefont{R.~K.}
  \bibnamefont{Schaefer}}, \bibinfo{journal}{Astrophys.J.}
  \textbf{\bibinfo{volume}{452}}, \bibinfo{pages}{495} (\bibinfo{year}{1995}),
  \eprint{astro-ph/9502087}.

\bibitem[{\citenamefont{Spergel and Steinhardt}(2000)}]{Spergel:1999mh}
\bibinfo{author}{\bibfnamefont{D.~N.} \bibnamefont{Spergel}} \bibnamefont{and}
  \bibinfo{author}{\bibfnamefont{P.~J.} \bibnamefont{Steinhardt}},
  \bibinfo{journal}{Phys.Rev.Lett.} \textbf{\bibinfo{volume}{84}},
  \bibinfo{pages}{3760} (\bibinfo{year}{2000}), \eprint{astro-ph/9909386}.

\bibitem[{\citenamefont{Wandelt et~al.}(2000)\citenamefont{Wandelt, Dave,
  Farrar, McGuire, Spergel et~al.}}]{Wandelt:2000ad}
\bibinfo{author}{\bibfnamefont{B.~D.} \bibnamefont{Wandelt}},
  \bibinfo{author}{\bibfnamefont{R.}~\bibnamefont{Dave}},
  \bibinfo{author}{\bibfnamefont{G.~R.} \bibnamefont{Farrar}},
  \bibinfo{author}{\bibfnamefont{P.~C.} \bibnamefont{McGuire}},
  \bibinfo{author}{\bibfnamefont{D.~N.} \bibnamefont{Spergel}},
  \bibnamefont{et~al.}, pp. \bibinfo{pages}{263--274} (\bibinfo{year}{2000}),
  \eprint{astro-ph/0006344}.

\bibitem[{\citenamefont{Firmani et~al.}(2000)\citenamefont{Firmani, D'Onghia,
  Avila-Reese, Chincarini, and Hernandez}}]{Firmani:2000ce}
\bibinfo{author}{\bibfnamefont{C.}~\bibnamefont{Firmani}},
  \bibinfo{author}{\bibfnamefont{E.}~\bibnamefont{D'Onghia}},
  \bibinfo{author}{\bibfnamefont{V.}~\bibnamefont{Avila-Reese}},
  \bibinfo{author}{\bibfnamefont{G.}~\bibnamefont{Chincarini}},
  \bibnamefont{and}
  \bibinfo{author}{\bibfnamefont{X.}~\bibnamefont{Hernandez}},
  \bibinfo{journal}{Mon.Not.Roy.Astron.Soc.} \textbf{\bibinfo{volume}{315}},
  \bibinfo{pages}{L29} (\bibinfo{year}{2000}), \eprint{astro-ph/0002376}.

\bibitem[{\citenamefont{Rocha et~al.}(2012)\citenamefont{Rocha, Peter, Bullock,
  Kaplinghat, Garrison-Kimmel et~al.}}]{Rocha:2012}
\bibinfo{author}{\bibfnamefont{M.}~\bibnamefont{Rocha}},
  \bibinfo{author}{\bibfnamefont{A.~H.~G.} \bibnamefont{Peter}},
  \bibinfo{author}{\bibfnamefont{J.~S.} \bibnamefont{Bullock}},
  \bibinfo{author}{\bibfnamefont{M.}~\bibnamefont{Kaplinghat}},
  \bibinfo{author}{\bibfnamefont{S.}~\bibnamefont{Garrison-Kimmel}},
  \bibnamefont{et~al.} (\bibinfo{year}{2012}), \eprint{1208.3025}.

\bibitem[{\citenamefont{Peter et~al.}(2012)\citenamefont{Peter, Rocha, Bullock,
  and Kaplinghat}}]{Peter:2012}
\bibinfo{author}{\bibfnamefont{A.~H.~G.} \bibnamefont{Peter}},
  \bibinfo{author}{\bibfnamefont{M.}~\bibnamefont{Rocha}},
  \bibinfo{author}{\bibfnamefont{J.~S.} \bibnamefont{Bullock}},
  \bibnamefont{and}
  \bibinfo{author}{\bibfnamefont{M.}~\bibnamefont{Kaplinghat}}
  (\bibinfo{year}{2012}), \eprint{1208.3026}.

\bibitem[{\citenamefont{Salucci et~al.}(2007)\citenamefont{Salucci, Lapi,
  Tonini, Gentile, Yegorova et~al.}}]{Salucci:2007tm}
\bibinfo{author}{\bibfnamefont{P.}~\bibnamefont{Salucci}},
  \bibinfo{author}{\bibfnamefont{A.}~\bibnamefont{Lapi}},
  \bibinfo{author}{\bibfnamefont{C.}~\bibnamefont{Tonini}},
  \bibinfo{author}{\bibfnamefont{G.}~\bibnamefont{Gentile}},
  \bibinfo{author}{\bibfnamefont{I.}~\bibnamefont{Yegorova}},
  \bibnamefont{et~al.}, \bibinfo{journal}{Mon.Not.Roy.Astron.Soc.}
  \textbf{\bibinfo{volume}{378}}, \bibinfo{pages}{41} (\bibinfo{year}{2007}),
  \eprint{astro-ph/0703115}.

\bibitem[{\citenamefont{Kaplinghat et~al.}(2000)\citenamefont{Kaplinghat, Knox,
  and Turner}}]{Kaplinghat:2000vt}
\bibinfo{author}{\bibfnamefont{M.}~\bibnamefont{Kaplinghat}},
  \bibinfo{author}{\bibfnamefont{L.}~\bibnamefont{Knox}}, \bibnamefont{and}
  \bibinfo{author}{\bibfnamefont{M.~S.} \bibnamefont{Turner}},
  \bibinfo{journal}{Phys.Rev.Lett.} \textbf{\bibinfo{volume}{85}},
  \bibinfo{pages}{3335} (\bibinfo{year}{2000}), \eprint{astro-ph/0005210}.

\bibitem[{\citenamefont{Casas et~al.}(1992)\citenamefont{Casas, Garcia-Bellido,
  and Quiros}}]{Casas:1991ky}
\bibinfo{author}{\bibfnamefont{J.}~\bibnamefont{Casas}},
  \bibinfo{author}{\bibfnamefont{J.}~\bibnamefont{Garcia-Bellido}},
  \bibnamefont{and} \bibinfo{author}{\bibfnamefont{M.}~\bibnamefont{Quiros}},
  \bibinfo{journal}{Class.Quant.Grav.} \textbf{\bibinfo{volume}{9}},
  \bibinfo{pages}{1371} (\bibinfo{year}{1992}), \eprint{hep-ph/9204213}.

\bibitem[{\citenamefont{Anderson and Carroll}(1997)}]{Anderson:1997un}
\bibinfo{author}{\bibfnamefont{G.~W.} \bibnamefont{Anderson}} \bibnamefont{and}
  \bibinfo{author}{\bibfnamefont{S.~M.} \bibnamefont{Carroll}}
  (\bibinfo{year}{1997}), \eprint{astro-ph/9711288}.

\bibitem[{\citenamefont{Amendola}(2000)}]{Amendola:1999er}
\bibinfo{author}{\bibfnamefont{L.}~\bibnamefont{Amendola}},
  \bibinfo{journal}{Phys.Rev.} \textbf{\bibinfo{volume}{D62}},
  \bibinfo{pages}{043511} (\bibinfo{year}{2000}), \eprint{astro-ph/9908023}.

\bibitem[{\citenamefont{Hoffman}(2003)}]{Hoffman:2003ru}
\bibinfo{author}{\bibfnamefont{M.~B.} \bibnamefont{Hoffman}}
  (\bibinfo{year}{2003}), \eprint{astro-ph/0307350}.

\bibitem[{\citenamefont{Farrar and Peebles}(2004)}]{Farrar:2003uw}
\bibinfo{author}{\bibfnamefont{G.~R.} \bibnamefont{Farrar}} \bibnamefont{and}
  \bibinfo{author}{\bibfnamefont{P.~J.~E.} \bibnamefont{Peebles}},
  \bibinfo{journal}{Astrophys.J.} \textbf{\bibinfo{volume}{604}},
  \bibinfo{pages}{1} (\bibinfo{year}{2004}), \eprint{astro-ph/0307316}.

\bibitem[{\citenamefont{Bean et~al.}(2008{\natexlab{a}})\citenamefont{Bean,
  Flanagan, and Trodden}}]{Bean:2007nx}
\bibinfo{author}{\bibfnamefont{R.}~\bibnamefont{Bean}},
  \bibinfo{author}{\bibfnamefont{E.~E.} \bibnamefont{Flanagan}},
  \bibnamefont{and} \bibinfo{author}{\bibfnamefont{M.}~\bibnamefont{Trodden}},
  \bibinfo{journal}{New J.Phys.} \textbf{\bibinfo{volume}{10}},
  \bibinfo{pages}{033006} (\bibinfo{year}{2008}{\natexlab{a}}),
  \eprint{0709.1124}.

\bibitem[{\citenamefont{Bean et~al.}(2008{\natexlab{b}})\citenamefont{Bean,
  Flanagan, and Trodden}}]{Bean:2007ny}
\bibinfo{author}{\bibfnamefont{R.}~\bibnamefont{Bean}},
  \bibinfo{author}{\bibfnamefont{E.~E.} \bibnamefont{Flanagan}},
  \bibnamefont{and} \bibinfo{author}{\bibfnamefont{M.}~\bibnamefont{Trodden}},
  \bibinfo{journal}{Phys.Rev.} \textbf{\bibinfo{volume}{D78}},
  \bibinfo{pages}{023009} (\bibinfo{year}{2008}{\natexlab{b}}),
  \eprint{0709.1128}.

\bibitem[{\citenamefont{Bean et~al.}(2008{\natexlab{c}})\citenamefont{Bean,
  Flanagan, Laszlo, and Trodden}}]{Bean:2008ac}
\bibinfo{author}{\bibfnamefont{R.}~\bibnamefont{Bean}},
  \bibinfo{author}{\bibfnamefont{E.~E.} \bibnamefont{Flanagan}},
  \bibinfo{author}{\bibfnamefont{I.}~\bibnamefont{Laszlo}}, \bibnamefont{and}
  \bibinfo{author}{\bibfnamefont{M.}~\bibnamefont{Trodden}},
  \bibinfo{journal}{Phys.Rev.} \textbf{\bibinfo{volume}{D78}},
  \bibinfo{pages}{123514} (\bibinfo{year}{2008}{\natexlab{c}}),
  \eprint{0808.1105}.

\bibitem[{\citenamefont{Corasaniti}(2008)}]{Corasaniti:2008kx}
\bibinfo{author}{\bibfnamefont{P.~S.} \bibnamefont{Corasaniti}},
  \bibinfo{journal}{Phys.Rev.} \textbf{\bibinfo{volume}{D78}},
  \bibinfo{pages}{083538} (\bibinfo{year}{2008}), \eprint{0808.1646}.

\bibitem[{\citenamefont{Cohen et~al.}(2008)\citenamefont{Cohen, Morrissey, and
  Pierce}}]{Cohen:2008nb}
\bibinfo{author}{\bibfnamefont{T.}~\bibnamefont{Cohen}},
  \bibinfo{author}{\bibfnamefont{D.~E.} \bibnamefont{Morrissey}},
  \bibnamefont{and} \bibinfo{author}{\bibfnamefont{A.}~\bibnamefont{Pierce}},
  \bibinfo{journal}{Phys.Rev.} \textbf{\bibinfo{volume}{D78}},
  \bibinfo{pages}{111701} (\bibinfo{year}{2008}), \eprint{0808.3994}.

\bibitem[{\citenamefont{Khoury and
  Weltman}(2004{\natexlab{a}})}]{Khoury:2003aq}
\bibinfo{author}{\bibfnamefont{J.}~\bibnamefont{Khoury}} \bibnamefont{and}
  \bibinfo{author}{\bibfnamefont{A.}~\bibnamefont{Weltman}},
  \bibinfo{journal}{Phys.Rev.Lett.} \textbf{\bibinfo{volume}{93}},
  \bibinfo{pages}{171104} (\bibinfo{year}{2004}{\natexlab{a}}),
  \eprint{astro-ph/0309300}.

\bibitem[{\citenamefont{Khoury and
  Weltman}(2004{\natexlab{b}})}]{Khoury:2003rn}
\bibinfo{author}{\bibfnamefont{J.}~\bibnamefont{Khoury}} \bibnamefont{and}
  \bibinfo{author}{\bibfnamefont{A.}~\bibnamefont{Weltman}},
  \bibinfo{journal}{Phys.Rev.} \textbf{\bibinfo{volume}{D69}},
  \bibinfo{pages}{044026} (\bibinfo{year}{2004}{\natexlab{b}}),
  \eprint{astro-ph/0309411}.

\bibitem[{\citenamefont{Brax et~al.}(2010)\citenamefont{Brax, van~de Bruck,
  Mota, Nunes, and Winther}}]{Brax:2010kv}
\bibinfo{author}{\bibfnamefont{P.}~\bibnamefont{Brax}},
  \bibinfo{author}{\bibfnamefont{C.}~\bibnamefont{van~de Bruck}},
  \bibinfo{author}{\bibfnamefont{D.~F.} \bibnamefont{Mota}},
  \bibinfo{author}{\bibfnamefont{N.~J.} \bibnamefont{Nunes}}, \bibnamefont{and}
  \bibinfo{author}{\bibfnamefont{H.~A.} \bibnamefont{Winther}},
  \bibinfo{journal}{Phys.Rev.} \textbf{\bibinfo{volume}{D82}},
  \bibinfo{pages}{083503} (\bibinfo{year}{2010}), \eprint{1006.2796}.

\bibitem[{\citenamefont{Gannouji et~al.}(2010)\citenamefont{Gannouji, Moraes,
  Mota, Polarski, Tsujikawa et~al.}}]{Gannouji:2010fc}
\bibinfo{author}{\bibfnamefont{R.}~\bibnamefont{Gannouji}},
  \bibinfo{author}{\bibfnamefont{B.}~\bibnamefont{Moraes}},
  \bibinfo{author}{\bibfnamefont{D.~F.} \bibnamefont{Mota}},
  \bibinfo{author}{\bibfnamefont{D.}~\bibnamefont{Polarski}},
  \bibinfo{author}{\bibfnamefont{S.}~\bibnamefont{Tsujikawa}},
  \bibnamefont{et~al.}, \bibinfo{journal}{Phys.Rev.}
  \textbf{\bibinfo{volume}{D82}}, \bibinfo{pages}{124006}
  (\bibinfo{year}{2010}), \eprint{1010.3769}.

\bibitem[{\citenamefont{Mota and Winther}(2011)}]{Mota:2010uy}
\bibinfo{author}{\bibfnamefont{D.~F.} \bibnamefont{Mota}} \bibnamefont{and}
  \bibinfo{author}{\bibfnamefont{H.~A.} \bibnamefont{Winther}},
  \bibinfo{journal}{Astrophys.J.} \textbf{\bibinfo{volume}{733}},
  \bibinfo{pages}{7} (\bibinfo{year}{2011}), \eprint{1010.5650}.

\bibitem[{\citenamefont{Nelson and Walsh}(2008)}]{Nelson:2008tn}
\bibinfo{author}{\bibfnamefont{A.}~\bibnamefont{Nelson}} \bibnamefont{and}
  \bibinfo{author}{\bibfnamefont{J.}~\bibnamefont{Walsh}},
  \bibinfo{journal}{Phys.Rev.} \textbf{\bibinfo{volume}{D77}},
  \bibinfo{pages}{095006} (\bibinfo{year}{2008}), \eprint{0802.0762}.

\bibitem[{\citenamefont{Feng et~al.}(2008)\citenamefont{Feng, Tu, and
  Yu}}]{Feng:2008mu}
\bibinfo{author}{\bibfnamefont{J.~L.} \bibnamefont{Feng}},
  \bibinfo{author}{\bibfnamefont{H.}~\bibnamefont{Tu}}, \bibnamefont{and}
  \bibinfo{author}{\bibfnamefont{H.-B.} \bibnamefont{Yu}},
  \bibinfo{journal}{JCAP} \textbf{\bibinfo{volume}{0810}}, \bibinfo{pages}{043}
  (\bibinfo{year}{2008}), \eprint{0808.2318}.

\bibitem[{\citenamefont{Cyburt et~al.}(2005)\citenamefont{Cyburt, Fields,
  Olive, and Skillman}}]{Cyburt:2004yc}
\bibinfo{author}{\bibfnamefont{R.~H.} \bibnamefont{Cyburt}},
  \bibinfo{author}{\bibfnamefont{B.~D.} \bibnamefont{Fields}},
  \bibinfo{author}{\bibfnamefont{K.~A.} \bibnamefont{Olive}}, \bibnamefont{and}
  \bibinfo{author}{\bibfnamefont{E.}~\bibnamefont{Skillman}},
  \bibinfo{journal}{Astropart.Phys.} \textbf{\bibinfo{volume}{23}},
  \bibinfo{pages}{313} (\bibinfo{year}{2005}), \eprint{astro-ph/0408033}.

\bibitem[{\citenamefont{Ackerman et~al.}(2009)\citenamefont{Ackerman, Buckley,
  Carroll, and Kamionkowski}}]{Ackerman:2008gi}
\bibinfo{author}{\bibfnamefont{L.}~\bibnamefont{Ackerman}},
  \bibinfo{author}{\bibfnamefont{M.~R.} \bibnamefont{Buckley}},
  \bibinfo{author}{\bibfnamefont{S.~M.} \bibnamefont{Carroll}},
  \bibnamefont{and}
  \bibinfo{author}{\bibfnamefont{M.}~\bibnamefont{Kamionkowski}},
  \bibinfo{journal}{Phys.Rev.} \textbf{\bibinfo{volume}{D79}},
  \bibinfo{pages}{023519} (\bibinfo{year}{2009}), \eprint{0810.5126}.

\bibitem[{\citenamefont{Hinshaw et~al.}(2009)}]{Hinshaw:2008kr}
\bibinfo{author}{\bibfnamefont{G.}~\bibnamefont{Hinshaw}} \bibnamefont{et~al.}
  (\bibinfo{collaboration}{WMAP Collaboration}),
  \bibinfo{journal}{Astrophys.J.Suppl.} \textbf{\bibinfo{volume}{180}},
  \bibinfo{pages}{225} (\bibinfo{year}{2009}), \eprint{0803.0732}.

\bibitem[{\citenamefont{Larson et~al.}(2011)\citenamefont{Larson, Dunkley,
  Hinshaw, Komatsu, Nolta et~al.}}]{Larson:2010gs}
\bibinfo{author}{\bibfnamefont{D.}~\bibnamefont{Larson}},
  \bibinfo{author}{\bibfnamefont{J.}~\bibnamefont{Dunkley}},
  \bibinfo{author}{\bibfnamefont{G.}~\bibnamefont{Hinshaw}},
  \bibinfo{author}{\bibfnamefont{E.}~\bibnamefont{Komatsu}},
  \bibinfo{author}{\bibfnamefont{M.}~\bibnamefont{Nolta}},
  \bibnamefont{et~al.}, \bibinfo{journal}{Astrophys.J.Suppl.}
  \textbf{\bibinfo{volume}{192}}, \bibinfo{pages}{16} (\bibinfo{year}{2011}),
  \eprint{1001.4635}.

\bibitem[{\citenamefont{Kolb and Turner}(1990)}]{K&T}
\bibinfo{author}{\bibfnamefont{E.~W.} \bibnamefont{Kolb}} \bibnamefont{and}
  \bibinfo{author}{\bibfnamefont{M.~S.} \bibnamefont{Turner}},
  \emph{\bibinfo{title}{The Early Universe}}, Frontiers in Physics
  (\bibinfo{publisher}{Addison-Wesley Publishing Company},
  \bibinfo{year}{1990}).

\bibitem[{\citenamefont{Steinhardt}(2005)}]{Steinhardt:2003iu}
\bibinfo{author}{\bibfnamefont{C.~L.} \bibnamefont{Steinhardt}},
  \bibinfo{journal}{Phys.Rev.} \textbf{\bibinfo{volume}{D71}},
  \bibinfo{pages}{043509} (\bibinfo{year}{2005}), \eprint{hep-ph/0308253}.

\bibitem[{\citenamefont{Birrell and Davies}(1986)}]{Birrell-GR}
\bibinfo{author}{\bibfnamefont{N.~D.} \bibnamefont{Birrell}} \bibnamefont{and}
  \bibinfo{author}{\bibfnamefont{P.~C.~W.} \bibnamefont{Davies}},
  \emph{\bibinfo{title}{Quantum Fields in Curved Space}}
  (\bibinfo{publisher}{Cambridge University Press}, \bibinfo{year}{1986}),
  \bibinfo{edition}{2nd} ed.

\bibitem[{\citenamefont{Brax et~al.}(2004)\citenamefont{Brax, van~de Bruck,
  Davis, Khoury, and Weltman}}]{Brax:2004qh}
\bibinfo{author}{\bibfnamefont{P.}~\bibnamefont{Brax}},
  \bibinfo{author}{\bibfnamefont{C.}~\bibnamefont{van~de Bruck}},
  \bibinfo{author}{\bibfnamefont{A.-C.} \bibnamefont{Davis}},
  \bibinfo{author}{\bibfnamefont{J.}~\bibnamefont{Khoury}}, \bibnamefont{and}
  \bibinfo{author}{\bibfnamefont{A.}~\bibnamefont{Weltman}},
  \bibinfo{journal}{Phys.Rev.} \textbf{\bibinfo{volume}{D70}},
  \bibinfo{pages}{123518} (\bibinfo{year}{2004}), \eprint{astro-ph/0408415}.

\bibitem[{\citenamefont{Edsjo and Gondolo}(1997)}]{Edsjo:1997bg}
\bibinfo{author}{\bibfnamefont{J.}~\bibnamefont{Edsjo}} \bibnamefont{and}
  \bibinfo{author}{\bibfnamefont{P.}~\bibnamefont{Gondolo}},
  \bibinfo{journal}{Phys.Rev.} \textbf{\bibinfo{volume}{D56}},
  \bibinfo{pages}{1879} (\bibinfo{year}{1997}), \eprint{hep-ph/9704361}.

\bibitem[{\citenamefont{Berestetskii et~al.}(1982)\citenamefont{Berestetskii,
  Lifshitz, and Pitaevskii}}]{Lifshitz-QED}
\bibinfo{author}{\bibfnamefont{V.}~\bibnamefont{Berestetskii}},
  \bibinfo{author}{\bibfnamefont{E.}~\bibnamefont{Lifshitz}}, \bibnamefont{and}
  \bibinfo{author}{\bibfnamefont{L.}~\bibnamefont{Pitaevskii}},
  \emph{\bibinfo{title}{Quantum Electrodynamics}} (\bibinfo{publisher}{Pergamon
  Press}, \bibinfo{year}{1982}), \bibinfo{edition}{2nd} ed.

\bibitem[{\citenamefont{Sommerfeld}(1931)}]{Sommerfeld:1931}
\bibinfo{author}{\bibfnamefont{A.}~\bibnamefont{Sommerfeld}},
  \bibinfo{journal}{Annalen der Physik} \textbf{\bibinfo{volume}{403}},
  \bibinfo{pages}{257} (\bibinfo{year}{1931}).

\bibitem[{\citenamefont{Arkani-Hamed et~al.}(2009)\citenamefont{Arkani-Hamed,
  Finkbeiner, Slatyer, and Weiner}}]{ArkaniHamed:2008qn}
\bibinfo{author}{\bibfnamefont{N.}~\bibnamefont{Arkani-Hamed}},
  \bibinfo{author}{\bibfnamefont{D.~P.} \bibnamefont{Finkbeiner}},
  \bibinfo{author}{\bibfnamefont{T.~R.} \bibnamefont{Slatyer}},
  \bibnamefont{and} \bibinfo{author}{\bibfnamefont{N.}~\bibnamefont{Weiner}},
  \bibinfo{journal}{Phys.Rev.} \textbf{\bibinfo{volume}{D79}},
  \bibinfo{pages}{015014} (\bibinfo{year}{2009}), \eprint{0810.0713}.

\bibitem[{\citenamefont{Lattanzi and Silk}(2009)}]{Lattanzi:2008qa}
\bibinfo{author}{\bibfnamefont{M.}~\bibnamefont{Lattanzi}} \bibnamefont{and}
  \bibinfo{author}{\bibfnamefont{J.~I.} \bibnamefont{Silk}},
  \bibinfo{journal}{Phys.Rev.} \textbf{\bibinfo{volume}{D79}},
  \bibinfo{pages}{083523} (\bibinfo{year}{2009}), \eprint{0812.0360}.

\bibitem[{\citenamefont{Robertson and Zentner}(2009)}]{Robertson:2009bh}
\bibinfo{author}{\bibfnamefont{B.}~\bibnamefont{Robertson}} \bibnamefont{and}
  \bibinfo{author}{\bibfnamefont{A.}~\bibnamefont{Zentner}},
  \bibinfo{journal}{Phys.Rev.} \textbf{\bibinfo{volume}{D79}},
  \bibinfo{pages}{083525} (\bibinfo{year}{2009}), \eprint{0902.0362}.

\bibitem[{\citenamefont{Buckley and Fox}(2010)}]{Buckley:2009in}
\bibinfo{author}{\bibfnamefont{M.~R.} \bibnamefont{Buckley}} \bibnamefont{and}
  \bibinfo{author}{\bibfnamefont{P.~J.} \bibnamefont{Fox}},
  \bibinfo{journal}{Phys.Rev.} \textbf{\bibinfo{volume}{D81}},
  \bibinfo{pages}{083522} (\bibinfo{year}{2010}), \eprint{0911.3898}.

\bibitem[{\citenamefont{Hannestad}(2000)}]{Hannestad:2000bs}
\bibinfo{author}{\bibfnamefont{S.}~\bibnamefont{Hannestad}}, pp.
  \bibinfo{pages}{81--88} (\bibinfo{year}{2000}), \eprint{astro-ph/0008422}.

\bibitem[{\citenamefont{'t~Hooft and Veltman}(1979)}]{tHooft:1978xw}
\bibinfo{author}{\bibfnamefont{G.}~\bibnamefont{'t~Hooft}} \bibnamefont{and}
  \bibinfo{author}{\bibfnamefont{M.}~\bibnamefont{Veltman}},
  \bibinfo{journal}{Nucl.Phys.} \textbf{\bibinfo{volume}{B153}},
  \bibinfo{pages}{365} (\bibinfo{year}{1979}).

\bibitem[{\citenamefont{Passarino and Veltman}(1979)}]{Passarino:1978jh}
\bibinfo{author}{\bibfnamefont{G.}~\bibnamefont{Passarino}} \bibnamefont{and}
  \bibinfo{author}{\bibfnamefont{M.}~\bibnamefont{Veltman}},
  \bibinfo{journal}{Nucl.Phys.} \textbf{\bibinfo{volume}{B160}},
  \bibinfo{pages}{151} (\bibinfo{year}{1979}).

\bibitem[{\citenamefont{Consoli}(1979)}]{Consoli:1979xw}
\bibinfo{author}{\bibfnamefont{M.}~\bibnamefont{Consoli}},
  \bibinfo{journal}{Nucl.Phys.} \textbf{\bibinfo{volume}{B160}},
  \bibinfo{pages}{208} (\bibinfo{year}{1979}).

\bibitem[{\citenamefont{Denner}(1993)}]{Denner:1991kt}
\bibinfo{author}{\bibfnamefont{A.}~\bibnamefont{Denner}},
  \bibinfo{journal}{Fortsch.Phys.} \textbf{\bibinfo{volume}{41}},
  \bibinfo{pages}{307} (\bibinfo{year}{1993}), \eprint{0709.1075}.

\bibitem[{\citenamefont{Gondolo et~al.}(2011)\citenamefont{Gondolo, Ko, and
  Omura}}]{Gondolo:2011eq}
\bibinfo{author}{\bibfnamefont{P.}~\bibnamefont{Gondolo}},
  \bibinfo{author}{\bibfnamefont{P.}~\bibnamefont{Ko}}, \bibnamefont{and}
  \bibinfo{author}{\bibfnamefont{Y.}~\bibnamefont{Omura}}
  (\bibinfo{year}{2011}), \eprint{1106.0885}.

\bibitem[{\citenamefont{Holdom}(1986)}]{Holdom:1985ag}
\bibinfo{author}{\bibfnamefont{B.}~\bibnamefont{Holdom}},
  \bibinfo{journal}{Phys.Lett.} \textbf{\bibinfo{volume}{B166}},
  \bibinfo{pages}{196} (\bibinfo{year}{1986}).

\bibitem[{\citenamefont{Feng et~al.}(2010)\citenamefont{Feng, Kaplinghat, and
  Yu}}]{Feng:2010zp}
\bibinfo{author}{\bibfnamefont{J.~L.} \bibnamefont{Feng}},
  \bibinfo{author}{\bibfnamefont{M.}~\bibnamefont{Kaplinghat}},
  \bibnamefont{and} \bibinfo{author}{\bibfnamefont{H.-B.} \bibnamefont{Yu}},
  \bibinfo{journal}{Phys.Rev.} \textbf{\bibinfo{volume}{D82}},
  \bibinfo{pages}{083525} (\bibinfo{year}{2010}), \eprint{1005.4678}.

\bibitem[{\citenamefont{Bjorken et~al.}(2009)\citenamefont{Bjorken, Essig,
  Schuster, and Toro}}]{Bjorken:2009mm}
\bibinfo{author}{\bibfnamefont{J.~D.} \bibnamefont{Bjorken}},
  \bibinfo{author}{\bibfnamefont{R.}~\bibnamefont{Essig}},
  \bibinfo{author}{\bibfnamefont{P.}~\bibnamefont{Schuster}}, \bibnamefont{and}
  \bibinfo{author}{\bibfnamefont{N.}~\bibnamefont{Toro}},
  \bibinfo{journal}{Phys.Rev.} \textbf{\bibinfo{volume}{D80}},
  \bibinfo{pages}{075018} (\bibinfo{year}{2009}), \eprint{0906.0580}.

\bibitem[{\citenamefont{Batell et~al.}(2009)\citenamefont{Batell, Pospelov, and
  Ritz}}]{Batell:2009yf}
\bibinfo{author}{\bibfnamefont{B.}~\bibnamefont{Batell}},
  \bibinfo{author}{\bibfnamefont{M.}~\bibnamefont{Pospelov}}, \bibnamefont{and}
  \bibinfo{author}{\bibfnamefont{A.}~\bibnamefont{Ritz}},
  \bibinfo{journal}{Phys.Rev.} \textbf{\bibinfo{volume}{D79}},
  \bibinfo{pages}{115008} (\bibinfo{year}{2009}), \eprint{0903.0363}.

\bibitem[{\citenamefont{Pospelov and Ritz}(2009)}]{Pospelov:2008jd}
\bibinfo{author}{\bibfnamefont{M.}~\bibnamefont{Pospelov}} \bibnamefont{and}
  \bibinfo{author}{\bibfnamefont{A.}~\bibnamefont{Ritz}},
  \bibinfo{journal}{Phys.Lett.} \textbf{\bibinfo{volume}{B671}},
  \bibinfo{pages}{391} (\bibinfo{year}{2009}), \eprint{0810.1502}.

\bibitem[{\citenamefont{Pospelov}(2009)}]{Pospelov:2008zw}
\bibinfo{author}{\bibfnamefont{M.}~\bibnamefont{Pospelov}},
  \bibinfo{journal}{Phys.Rev.} \textbf{\bibinfo{volume}{D80}},
  \bibinfo{pages}{095002} (\bibinfo{year}{2009}), \eprint{0811.1030}.

\bibitem[{\citenamefont{Binosi et~al.}(2009)\citenamefont{Binosi, Collins,
  Kaufhold, and Theussl}}]{JaxoDraw-v2}
\bibinfo{author}{\bibfnamefont{D.}~\bibnamefont{Binosi}},
  \bibinfo{author}{\bibfnamefont{J.}~\bibnamefont{Collins}},
  \bibinfo{author}{\bibfnamefont{C.}~\bibnamefont{Kaufhold}}, \bibnamefont{and}
  \bibinfo{author}{\bibfnamefont{L.}~\bibnamefont{Theussl}},
  \bibinfo{journal}{Comput.Phys.Commun.} \textbf{\bibinfo{volume}{180}},
  \bibinfo{pages}{1709} (\bibinfo{year}{2009}), \eprint{0811.4113}.

\bibitem[{\citenamefont{Binosi and Theussl}(2004)}]{JaxoDraw-v1}
\bibinfo{author}{\bibfnamefont{D.}~\bibnamefont{Binosi}} \bibnamefont{and}
  \bibinfo{author}{\bibfnamefont{L.}~\bibnamefont{Theussl}},
  \bibinfo{journal}{Comput.Phys.Commun.} \textbf{\bibinfo{volume}{161}},
  \bibinfo{pages}{76} (\bibinfo{year}{2004}), \eprint{hep-ph/0309015}.

\bibitem[{\citenamefont{Mertig et~al.}(1991)\citenamefont{Mertig, Bohm, and
  Denner}}]{FeynCalc}
\bibinfo{author}{\bibfnamefont{R.}~\bibnamefont{Mertig}},
  \bibinfo{author}{\bibfnamefont{M.}~\bibnamefont{Bohm}}, \bibnamefont{and}
  \bibinfo{author}{\bibfnamefont{A.}~\bibnamefont{Denner}},
  \bibinfo{journal}{Comput.Phys.Commun.} \textbf{\bibinfo{volume}{64}},
  \bibinfo{pages}{345} (\bibinfo{year}{1991}).

\end{thebibliography}

\end{document}